\documentclass[11pt,twoside]{atmp}

\usepackage{amsmath,amssymb,amsfonts,amscd,epsfig}

\usepackage{comment}
\usepackage{color}
\usepackage{graphicx}
\newcommand{\be}{\begin{equation}}
\newcommand{\beq}{\begin{equation}}
\newcommand{\eeq}{\end{equation}}
\newcommand{\ee}{\end{equation}}
\newcommand{\bea}{\begin{eqnarray}}
\newcommand{\eea}{\end{eqnarray}}
\newcommand{\ba}{\begin{array}}
\newcommand{\ea}{\end{array}}

\newcommand{\ga}{\gamma}
\newcommand\IZ{\mathbb{Z}}
\newcommand\IR{\mathbb{R}}
\newcommand{\IC}{\mathbb{C}}
\newcommand\IT{\mathbb{T}}
\newcommand{\II}{\mathbb{I}}

\def\rref#1{(\ref{#1})}

\begin{document}

\title{Theory of intersecting loops on a torus}

%\titlerunning{}        % if too long for running head

\author[J. E. Nelson and R. F. Picken]{J. E. Nelson and R. F. Picken}

\address{Dipartimento di Fisica, Sezione Teorica, Universit\`a degli Studi di
Torino\\ 
and INFN, Sezione di Torino\\ 
via Pietro Giuria 1, 10125 Torino, Italy\\}
\addressemail{nelson@to.infn.it}
\address{Departamento de Matem\'{a}tica and CAMGSD\\ - Centro de An\'{a}lise
Matem\'{a}tica, Geometria e Sistemas Din\^{a}micos\\
Instituto Superior T\'{e}cnico, Universidade de Lisboa\\
Avenida Rovisco Pais, 1049-001 Lisboa, Portugal\\}  
%lines should be separated with double backslashes: \\
\addressemail{rpicken@math.ist.utl.pt}


\begin{abstract}

We continue our investigation into intersections of closed paths on a
torus, to further our understanding of the commutator algebra of Wilson 
loop observables in $2+1$ quantum gravity, when the cosmological constant is 
negative. We give a concise review of previous results, e.g. that signed area 
phases relate observables assigned to homotopic loops, and present new 
developments in this theory of intersecting loops on a torus. We state 
precise rules to be applied at intersections of both straight and 
crooked/rerouted paths in the covering space ${\IR}^2$. Two concrete examples 
of combinations of different rules are presented.
 
%\keywords{Wilson observables \and $2+1$ quantum gravity \and Goldman bracket
%\and area phases} 
%\PACS{02.40.Gh \and 04.60.Kz }
%\subclass{81R50 \and 83C45 }
\end{abstract}

\maketitle

\section{Introduction} \label{intro} \noindent 

Quantum gravity in $2+1$ spacetime dimensions can be understood as a Chern-Simons 
theory, with structure group depending on the cosmological constant $\Lambda$ 
\cite{ach,wit}. Our interest arose from an approach started by Regge and one of us 
\cite{NR1,NRZ} based on quantizing the algebra of Wilson loops (traced holonomies) 
when the spatial manifold is a Riemann surface and $\Lambda < 0$. In this case  
the gauge group is $\hbox{SO}(2,2)$ or its spinor group is 
$\hbox{SL}(2,\IR)\otimes \hbox{SL}(2,\IR)$. The corresponding Poisson algebra, which is 
subsequently quantized, determines the bracket between two Wilson loops, through their 
intersections, thus making contact with the Goldman Poisson bracket \cite{gol} for 
loops on a surface. For a general genus surface including punctures, i.e. with 
boundary, one approach \cite{Czech} constructs these observables as the lengths of closed 
geodesics on a Riemann surface, using the Teichm\"uller space coordinates and the graph 
technique of Penner \cite{Penner} and Fock \cite{Fock1}. A partial comparison between these 
two approaches was outlined in \cite{CNR}. Another approach, combinatorial quantization,
parametrizes holonomies assigned to the edges of a fixed graph on the surface, to get
a finite-dimensional phase space with a non-linear constraint, which 
is then quantized using r - matrix methods \cite{cq1}-\cite{cq5}. 
  
When $\Lambda < 0$ and the spatial manifold is a torus $\IT^2$, in a series of articles 
\cite{NP1} - \cite{npint} the present authors used piecewise linear (PL) paths in the covering 
space ${\IR}^2$ to represent loops on the torus. The corresponding quantum holonomies are 
represented by pairs of $\hbox{SL}(2,\IR)$ matrices with non - commuting 
components, in the sense that components of matrices representing different holonomies 
i.e. arising from non - homotopic loops, may not commute. The holonomies arise from a quantum 
connection with constant non-commuting components, following \cite{mikpic}. This leads to the associated quantum 
curvature being non-zero to order $\hbar$ \cite{npint}.

The traces of these quantum holonomies satisfy commutators, considered as a quantum 
version of the Goldman bracket i.e. the Poisson brackets of observables for $2+1$ gravity
\cite{NP1} - \cite{NP}. In this approach the number of loop homotopy classes is not 
fixed, and indeed the quantum connection distinguishes the holonomy even along homotopic 
loops. Nonetheless, it is possible to get a consistent quantization picture amongst this 
collection of Wilson variables for different loops. 

Integrating over closed paths (loops) on the torus, non - commuting quantum connections $\hat{A}$ 
give rise to quantum holonomies $\hat U$, represented by $\hbox{SL}(2,\IR)$ quantum matrices, 
one for each holonomy (i.e. we only consider one matrix from each pair. The others can be treated 
similarly):
\be
\hat U_i = \exp \int_{\ga_i} \hat A,  \quad i=1,2,
\label{hol1}
\ee
where the cycles $\ga_i$, $i=1,2$ that generate the fundamental group $\pi_1(\IT^2)$ satisfy
\be
\ga_1^{\vphantom{-1}}\cdot\ga_2^{\vphantom{-1}}\cdot\ga_1^{-1}\cdot\ga_2^{-1} = {\II}.
\label{gp}
\ee

The non-commutativity of the quantum connections $\hat A$ implies, from \rref{hol1} the non-commutativity 
of the holonomies $\hat U_i$, since the quantum matrices that represent them satisfy {\it by both 
matrix and operator multiplication}, the $q$--commutation relation (see e.g. \cite{npint})
\beq
\hat U_1 \hat U_2  = q \hat U_2 \hat U_1, 
\label{fund2}
\eeq
where the $q$ parameter\footnote{Note that in $q$ the exponent of 
$\exp$ is dimensionless, when all physical constants are taken into account.} is
$q=\exp (- \frac {i \hbar \sqrt{-\Lambda}}{4})$  i.e. the matrices $\hat U_1,
\hat U_2$ form a matrix--valued Weyl pair. Relation \rref{fund2} is a special case of an area phase 
relation discussed in Section \ref{sub22}.

An alternative expression of equation \rref{fund2} is
\beq
\hat U_1 \hat U_2  {\hat U_1}^{-1} {\hat U_2}^{-1} = q 
\label{fund3}
\eeq
and it is clear that equation \rref{fund3} can be understood as a $q$-deformed representation of equation 
\rref{gp}. 

The plan of the paper is as follows. In Section \ref{sec2} we review some interesting
features of the quantum geometry that has emerged. These are the
representation of loops on the torus by PL paths between
integer points in $\IR^2$ and 
%constant matrix--valued connections applied to a much
%larger class of loops, and a definition of a $q$--deformed representation of the
%fundamental group where 
signed area phases which relate the quantum matrices assigned
to homotopic loops. In Section \ref{sec3} intersections and reroutings, and two
quantizations of the Poisson bracket between paths are described: the `direct'
quantization and the `refined' quantization. We also discuss the quantum nature
of intersections (expressed as commutators) of straight paths (i.e. straight in
$\IR^2$) and `crooked' paths (those resulting from previous reroutings).
% by using the concepts of integer points and relative phases for a crooked rerouting. 
Section \ref{sec4} is completely new. We describe some new features of the theory of
intersecting loops on a torus, and give precise rules to be applied at
intersections of both straight and crooked paths which guarantee the reproduction of 
the corresponding straight - straight path result. We present two concrete examples of 
combinations of different rules. In Section \ref{concl} we present our conclusions.

\section{Piecewise linear paths and quantum holonomy matrices}\label{sec2}
\subsection{Piecewise linear paths}\label{sub21}
\begin{figure}
\begin{center}
\includegraphics[width=16pc]{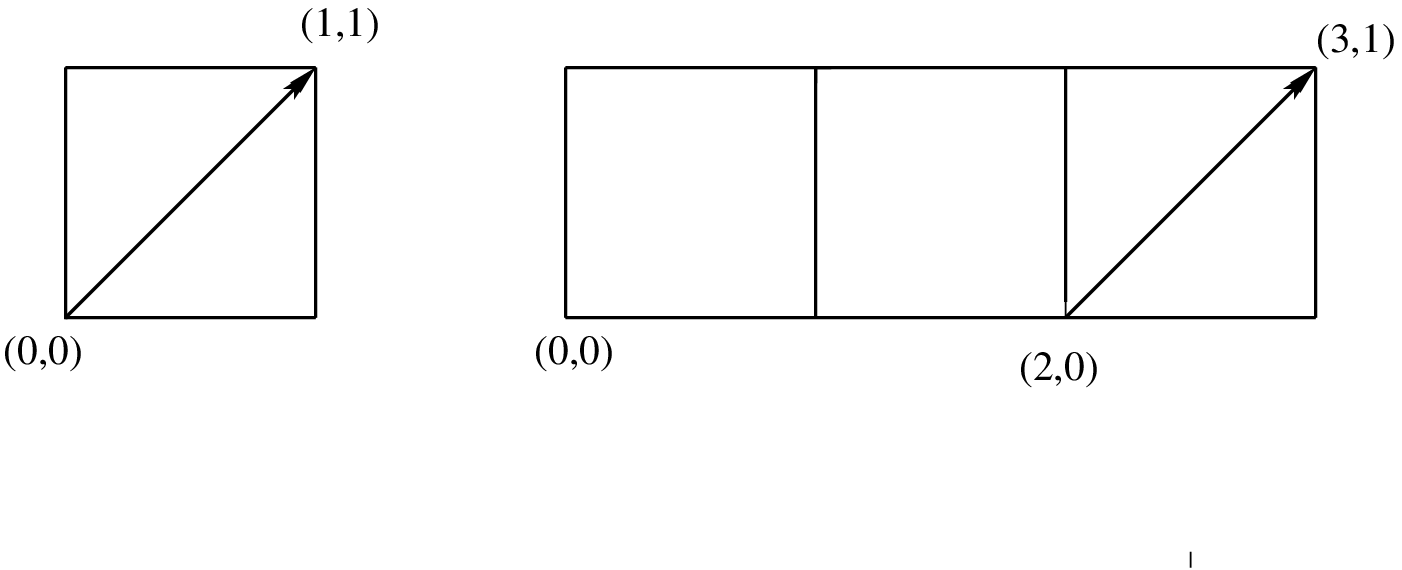}
\end{center}
\caption{\label{fund} The path from $O$ = $(0,0)$ to $(1,1)$ and the path from
$(2,0)$ to $(3,1)$ represent the same loop on the torus.}
\end{figure}

We will identify loops (closed paths) on the torus $\IT^2=\IR^2/\IZ^2$  
with paths on its covering space ${\IR}^2$, i.e. we represent all loops on the torus
by PL paths on ${\IR}^2$ between integer points $(m,n)\in
\IZ^2$. All these integer points are identified, and correspond to the same
point on the torus. A path in ${\IR}^2$ representing a loop on the torus can
therefore be replaced by any parallel path starting at a different integer
point, e.g. the path from $O=(0,0)$ to $(1,1)$ represents the same loop as the
path from $(2,0)$ to $(3,1)$, as shown in Figure \ref{fund}.

A natural subclass of paths in ${\IR}^2$ are those straight paths denoted
$p=(m,n)$ that start at the origin $O =(0,0)$ and end at an integer point
$(m,n)\in \IZ^2$. They generalise the cycles $\gamma_1,\, \gamma_2$
(corresponding to the paths $(1,0)$ and $(0,1)$ respectively). When the path
$(m,n)$ is a multiple of another integer path, i.e. $(m,n)=c(m',n')$, where
$m,n,c,m',n'$ are all integers, with $c\geq 2$, we say it is {\it reducible}.
Otherwise it is irreducible. 

The identification between loops on the torus $\IT^2=\IR^2/\IZ^2$ and PL paths
in $\IR^2$ can be  further understood by considering the concept of fundamental
reduction. In \cite{goldman} we introduced this concept in
order to better study the intersections between paths $p_1$ and $p_2$. It
consists of reducing one or more paths to a fundamental domain of 
$\IR^2$, namely the unit square with vertices $(0,0), (1,0)$ and $(1,1), (0,1)$, as 
follows: a path that passes through more than one cell (a unit
square in $\IR^2$) consists of ordered segments, each of which passes through
only one cell. The fundamental reduction is obtained by superimposing each of
these cells, in the order of the segments, on the fundamental domain (the
square, or cell, with vertices $(0,0), (1,0)$ and $(1,1), (0,1)$). In this
fundamental domain the left and right edges should be identified, and similarly
for the top and bottom edges. 

\begin{figure}[hbtp]
\centering
\includegraphics[height=2cm]{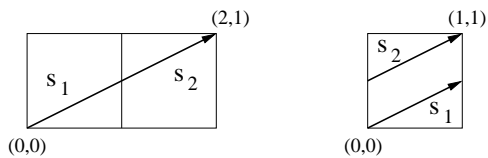}
%{21a.eps}
%\hspace{2cm}
%\includegraphics[height=2cm]{21br.eps}
\caption{The path $(2,1)$ and its fundamental reduction.}
\label{21}
\end{figure}
\noindent 

\begin{figure}[hbtp]
\centering
\includegraphics[height=3cm]{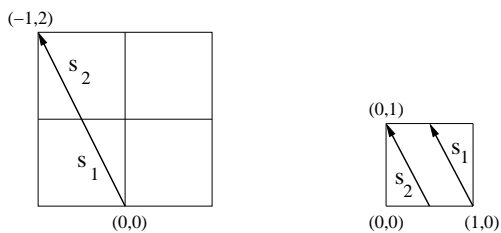}
%\includegraphics[width=5cm]{12j.eps}
%{12a.eps}
%\hspace{2cm}
%\includegraphics[width=2cm]{12b.eps}
\caption{The path $(-1,2)$ and its fundamental reduction.}
\label{-12}
\end{figure}

Two examples of fundamental reduction for straight paths are shown in Figures
\ref{21} and \ref{-12}. Figure \ref{21} shows a straight path in the first
quadrant, namely the path $(2,1)$, with its two segments labelled $s_1$ and
$s_2$ (in that order), and its reduction to the fundamental domain, 
whereas Figure \ref{-12} shows a straight path in the second quadrant,
namely $(-1,2)$, and its two segments $s_1$ and $s_2$ (in that order). Note
that, when fundamentally reduced, this path starts at $(1,0)$ (not the origin
$O=(0,0)$) and ends at $(0,1)$. This is because its first cell does not coincide
with the fundamental domain, and the same is true for non--reduced paths which start 
in all quadrants except the first. 
  
It is important that in the fundamental domain diagrams (the diagrams on the
right) of Figures \ref{21} and \ref{-12} the left and right edges should be
identified, and similarly for the top and bottom edges, i.e. these diagrams
depict a loop on the torus.

\subsection{Quantum holonomy matrices and signed area}\label{sub22}

As in \rref{hol1} a quantum matrix is assigned to any straight path $(m.n)$ by
\be
\hat{U}_{(m,n)}= \exp \int_{(m,n)} \hat{A} 
\label{Umn}
\ee
and this clearly extends straightforwardly to any PL path between integer points by
assigning a quantum matrix to each linear segment of the path, as in \rref{Umn},
and multiplying the matrices in the same order as the segments along the path.
This prescription obviously coincides with the general relation:
\be
p\rightarrow \hat{U}_p = {P }\exp \int_{p} \hat{A}.
\label{hol}
\ee
where $ P$ denotes path-ordering.

\begin{figure}
\begin{center}
\includegraphics[width=16pc]{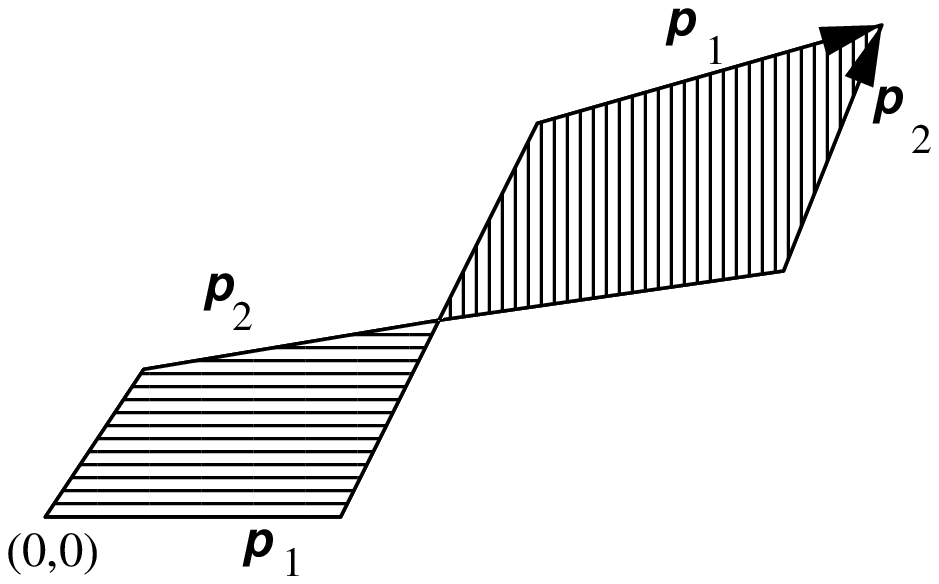}
\end{center}
\caption{\label{signedarea} The signed area between $p_1$ and $p_2$ is the area
of the horizontally shaded region minus the area of the vertically shaded
region.}
\end{figure}

In the covering space ${\IR}^2$, two homotopic loops on the torus are
represented by two PL paths on the plane, $p_1,\, p_2$, with the same integer
starting point and the same integer endpoint.  It was shown in \cite{goldman}
that the following relationship holds for the respective quantum matrices:
\begin{equation}
\hat{U}_{p_1}=q^{S(p_1,p_2)}\hat{U}_{p_2},
\label{areaphase}
\end{equation}
where $S(p_1,p_2)$ denotes the signed area enclosed between the paths
$p_1$ and $p_2$. Equation \rref {areaphase} generalises equation \rref{fund2},
for which $S(p_1,p_2)$ is the area of the square with vertices $(0,0), \, (0,1),
\, (1,0)$ and $(1,1)$, i.e. 1. For the general case, the signed area between two
PL paths is defined as follows: for any finite region $R$ enclosed by $p_1$ and
$p_2$, if the boundary of $R$ consists of oriented segments of $p_1$ and
$p_2^{-1}$ (the path $p_2$ followed in the opposite direction), and is globally
oriented in the positive (anticlockwise), or negative (clockwise) sense, this
gives a contribution of $+{\rm area}(R)$ , or $-{\rm area}(R)$ respectively, to
the signed sum $S(p_1,p_2)$ (otherwise the contribution is zero). See Figure
\ref{signedarea}.

As remarked in \cite{npint}, the signed area phases are highly suggestive 
of an underlying geometry with (abelian) surface transports, together with the more 
conventional parallel transports along paths - see the discussion on \mbox{$2$ - dimensional} 
holonomy in \cite{FMP}. For this we need two Lie groups, $G$ and $E$ (related to each other 
to form a crossed module of groups), and connection \mbox{$1$ -} and \mbox{$2$ - forms} with values in the 
Lie algebras of $G$ and $E$ respectively. The \mbox{$1$ - form} connection should obviously be 
$\hat{A}$, and the natural candidate for the connection \mbox{$2$ - form} is the 
non-vanishing quantum curvature \mbox{$2$ - form} of $\hat{A}$ (see \cite{npint}). Since we 
are dealing with quantum forms, the geometric theory of \cite{FMP} cannot be applied directly, 
but it certainly requires investigation.

\section{Intersecting paths and commutators}\label{sec3}

That Wilson loops associated to intersecting paths on surfaces have non--zero
Poisson brackets was noted in \cite{NRZ}. These are related to the Goldman
bracket \cite{gol} for the traces $T(\gamma)= {\rm tr}\, U_\ga$, $U_\ga \in
SL(2,\IR)$, defined on homotopy classes of loops on a surface
\be
\{T(\ga_1), T(\ga_2)\} = \sum_{I \in \ga_1 \sharp \ga_2}
\epsilon(\ga_1,\ga_2,I)(T(\ga_1I\ga_2) -
T(\ga_1I\ga_2^{-1})).
\label{gold}
\ee
Here $\ga_1 \sharp \ga_2$ denotes the set of transversal intersection points of $\ga_1$ and 
$\ga_2$, i.e. their tangent directions at their intersection point $I$ do not coincide. 
The intersection number of $\ga_1$ and $\ga_2$ at $I$ is denoted $\epsilon(\ga_1,\ga_2,I)$, and 
$\ga_1I\ga_2$, $\ga_1I\ga_2^{-1}$ are the loops rerouted at $I$ along $\ga_2$ or $\ga_2^{-1}$. 
We call them positive and negative reroutings, respectively. In this article we only discuss the positive 
reroutings, the negative reroutings can be treated similarly.
See below for a more detailed description of reroutings. 

Using the PL description of Section \ref{sub21}, for straight paths $p_1=(m,n)$ and
$p_2=(s,t)$ this takes the form:
\be 
\left\{T(m,n), T(s,t)\right\} = (mt-ns)
(T(m+s,n+t)- T(m-s,n-t)). 
\label{straightforward}
\ee 
Here $mt-ns$ is the total intersection number (see \rref{int}). In
\cite{goldman} it was shown  that the Wilson loops $\hat{T}(p)$ satisfy:
\begin{equation}
[\hat{T}(m,n), \hat{T}(s,t)]=
(q^{\frac{mt-ns}{2}}-q^{-\frac{mt-ns}{2}}) \left(\hat{T}(m+s,n+t) -
\hat{T}(m-s,n-t)\right)
\label{qgb1}
\end{equation}
i.e. a direct quantization of \rref{straightforward}, with the total
intersection number $mt-ns$ replaced by a quantum total intersection number (the
first factor on the right hand side (r.h.s.) of \rref{qgb1}). 

A refined but equivalent quantization was also obtained in \cite{goldman}, where
each rerouting appears as a separate term, and the relative area phases of these
different but homotopic reroutings are taken into acccount:
\be
[\hat{T}(p_1), \hat{T}(p_2)] = \sum_{I\in p_1 \sharp p_2}
(q^{\epsilon(p_1,p_2,I)} - 1)\hat{T}(p_1Ip_2)  
+ (q^{-\epsilon(p_1,p_2,I)} - 1)\hat{T}(p_1Ip_2^{-1})
\label{qgold}
\ee
which quantizes the bracket \rref{gold} (with loops $\ga$ substituted by paths
$p$) by replacing the intersection numbers $\epsilon(p_1,p_2,I)$ by quantum
intersection numbers $(q^{\epsilon(p_1,p_2,I)} - 1)$. 

Note that in \rref{qgold} paths which are {\it rerouted} at the intersection
points $I$ appear on the r.h.s. There are a number of important features of
reroutings which derive from intersections, whch we briefly summarize.

The intersection number between two paths $p_1$ and $p_2$, corresponding to
loops on the torus, at an intersection point $I$, is defined to be (for single,
not multiple, intersections) $+1$ if the angle between the tangent vector of
$p_1$ at $S$ and the tangent vector of $p_2$ at $S$ is between $0$ and $180$
degrees, and $-1$ if it is between $180$ and $360$ degrees.

A rerouted path is denoted $p_1Ip_2$, or $p_1Ip_2^{-1}$, and is the path that
follows $p_1$ as far as the intersection point $I$, then follows $p_2$ (or the
inverse loop $p_2^{-1}$) from $I$ back to $I$, and finally proceeds along $p_1$
from $I$ back to the starting point of $p_1$. This can be thought of as
`inserting' the path $p_2$ or $p_2^{-1}$ into the path $p_1$ at the intersection point $I$. 
Note that the point $I$ may occur at the origin of $\IR^2$, in which case we
follow $p_2$ straight away. 

\begin{figure}
\begin{center}
\includegraphics[width=3cm]{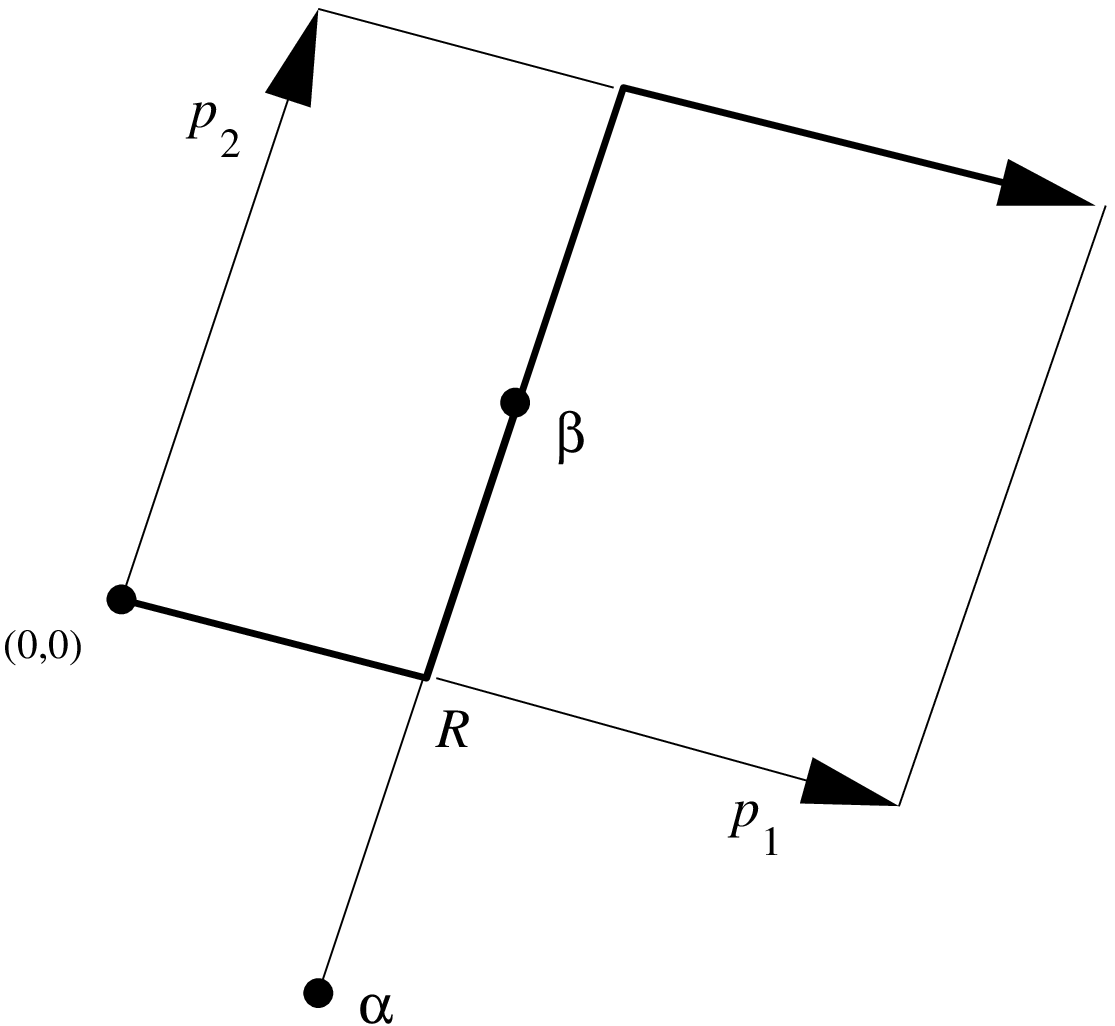}
\hspace{1cm}
\includegraphics[width=3cm]{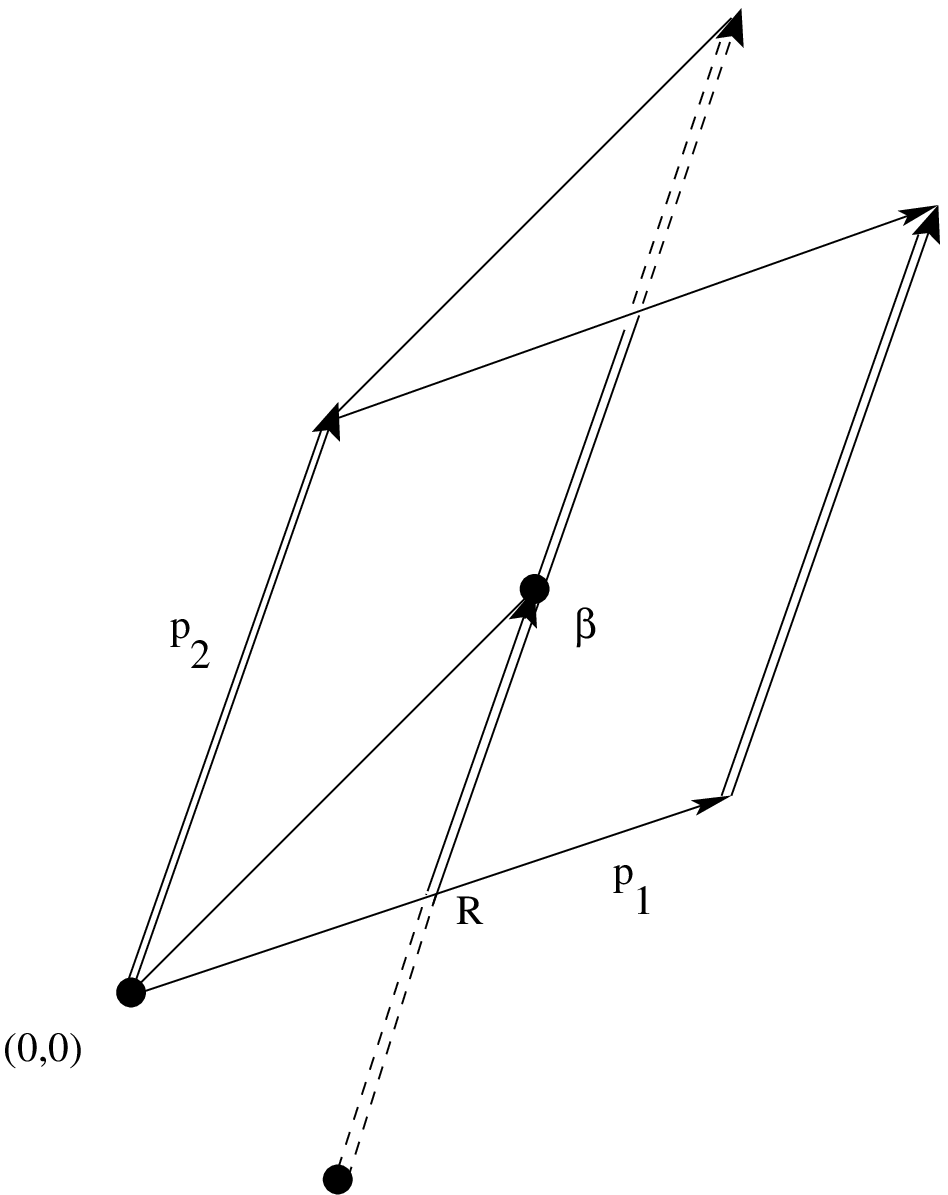}
\end{center}
\caption{\label{rerouting} The parallelogram with edges $p_1$ and $p_2$, and the
rerouting $p_1Rp_2$ displayed as dark segments. The segment between integer
points $\alpha$ and $\beta$ is a parallel copy of $p_2$, translated from
starting at the origin to starting at $\alpha$; the grid of all other integer
points is not displayed. The figure on the right shows an analogous situation with a 
crooked $p_2$ represented by  a double line.}
\end{figure}

Here we shall work directly in $\IR^2$. Even though all integer points are the
same when projected down to the torus, a very clear picture of where the
intersection point $R$ is located along both paths is obtained by fixing $p_1$
and parallel translating $p_2$ to start at a new integer point, denoted
$\alpha$, in such a way that it intersects $p_1$ at $R$, as shown in Figure
\ref{rerouting}. The path $p_2$, inserted into $p_1$, may
be straight (the left figure of Figure \ref{rerouting}) or crooked (the right figure).

It is therefore clear why intersections occur and how they give rise to
reroutings, i.e. for an intersection to occur, it is necessary that either
$p_2^\alpha$ (the path $p_2$ parallel translated to start at $\alpha$)
intersects $p_1$ in a point $R$ (which may be the origin, but not the endpoint
of $p_1$), or equivalently, the endpoint $\beta$ is such that the appropriate
$p_2^\alpha$ ending in $\beta$ intersects $p_1$ in a point $R$ as before.

\begin{figure}
\begin{center}
\includegraphics[width=12pc]{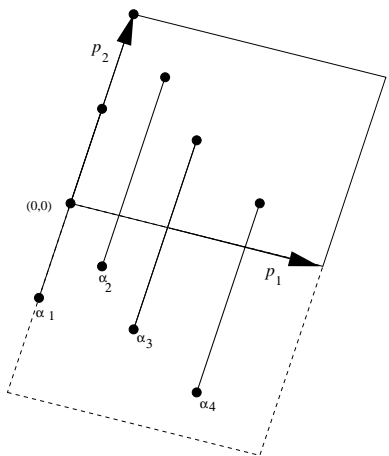}
\end{center}
\caption{\label{preparallelogram} The parallelogram and pre - parallelogram (dashed line) 
for $p_1$ and $p_2$, showing some of the integer points (black dots) and the corresponding 
parallel - translated copies of $p_2$.}
\end{figure}

When $p_2$ is straight, the possible starting points $\alpha$ are the integer 
points lying in a `pre -parallelogram' shown in Figure \ref{preparallelogram}, and 
the total intersection number (counting multiplicities) is therefore 
\be 
\epsilon (p_1,p_2) = \det \left(p_1, p_2\right).
\label{int}
\ee
i.e. in geometric terms, the area of the parallelogram or the pre - parallelogram.
This is also guaranteed by Pick's theorem \cite{pick} for the area $A(P)$ of a
planar polygon $P$ with
vertices at integer points of the plane 
\be
A(P) = I(P) + \frac{B(P)}{2} -1.
\label{pick}\ee
where $I(P)$ is the number of interior integer points and $B(P)$ is the number
of 
boundary integer points, i.e. $A(P)$ is the number of integer starting points
$\alpha$ 
in the pre - parallelogram, or equivalently the number of integer end points
$\beta$ in the parallelogram. 

When the second path $p_2$ in the rerouting $p_1Rp_2$ is crooked, and the first
path $p_1$ is straight, as on the r.h.s. of Figure \ref{rerouting}, unexpected
extra phases occur, i.e. there is a discrepancy when the inserted path is
crooked. In \cite{npint} we derived a formula for the relative phase of such
reroutings compared to the `first' rerouting, i.e. the rerouting that occurs
straight away at the origin (the origin is always an intersection point for any
two paths). The result is that the signed area, or relative phase, between the
rerouting $p_1Rp_2$ and the rerouting at the origin $p_1Op_2$ is
\be
 S(p_1Rp_2, p_1Op_2) = \det (\beta, \overline{p_2})
\label{s3}\ee
where $\beta$ is the integer endpoint associated to the intersection point $R$,
and $\overline{p_2}$ denotes the integer endpoint of $p_2$. Note that in \rref{s3} 
we could equally well use the integer starting point $\alpha$ instead of $\beta$, since 
$\beta - \alpha = \overline{p_2}$. This will be a key result for Section \ref{sub43}. 

%\subsection{The relative phase for a crooked rerouting}\label{sub41}

\section{Extending the refined bracket to general loops}\label{sec4}
%\blue{rearrangement putting rules and examples together, genrefr15}
%\red{changes and insertions in this section highlighted in red}

In \cite{npint} the refined formula for the Goldman bracket \rref{qgold} was
shown to be equivalent to the direct quantization \rref{qgb1}, but only for 
straight paths on the l.h.s., whereas on the r.h.s. the terms correspond to 
crooked or rerouted paths, as mentioned in Section \ref{sec3}. Here we show how the refined 
formula can be
extended to crooked paths $p_1,\, p_2$ on the l.h.s., in a way that is
consistent with the area phases.

Having crooked paths in the bracket brings in new features, e.g. the
intersection points between the paths $p_1$ and $p_2$ need not all have the same
intersection number, unlike in the case of two straight paths. Also, it could
happen that an intersection occurs at a point where a crooked path changes
direction. Finally intersections between $p_1$ and parallel translated copies of
$p_2$ may occur which would not be there if the paths were replaced by their
straightened versions $\overline{p_i}$. These new situations should be handled
by introducing additional rules which we now discuss.

\subsection{Rule one - the V-intersection rule}\label{sub41}
 
%\red{In this subsection we want to have more discussion about the symmetric
%intersection numbers that appear in the unrefined formula for straight-straight
%and the non-symmetric intersection numbers that appear in the refined formula
%for straight-straight.}

\begin{figure}[hbpt]
\centering
\includegraphics[height=2cm]{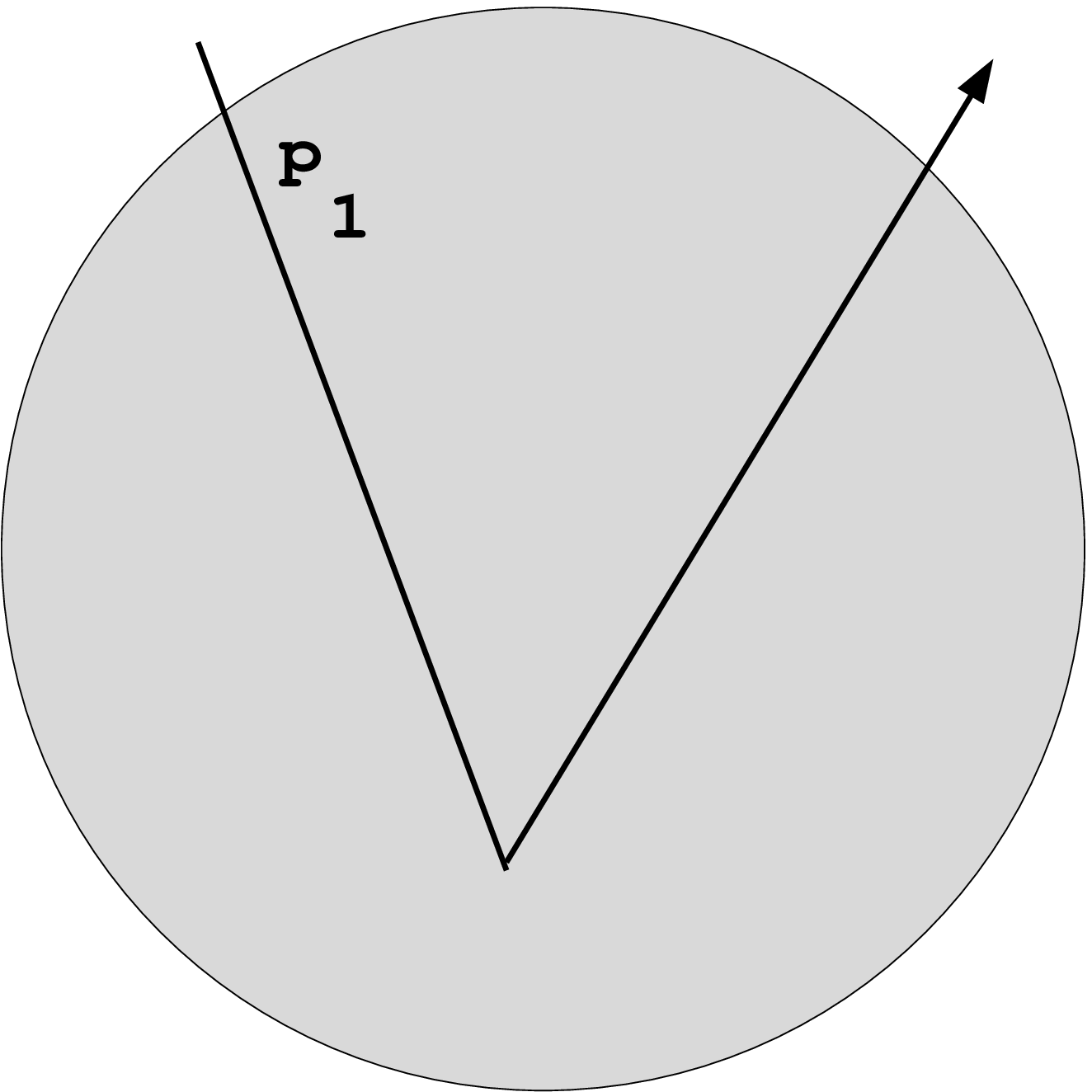}
\hspace{1cm}
\includegraphics[height=2cm]{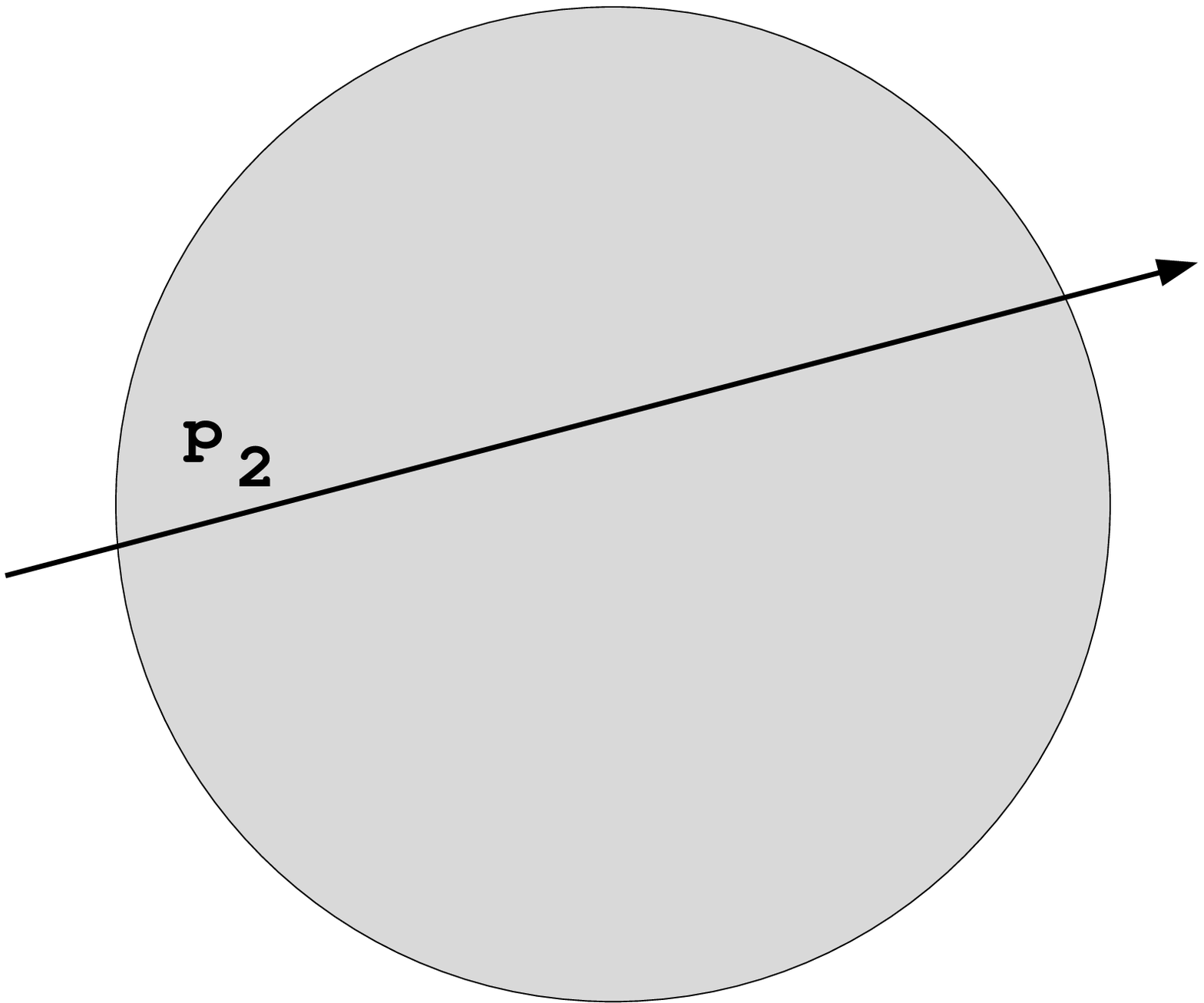}
\hspace{1cm}
\includegraphics[height=2cm]{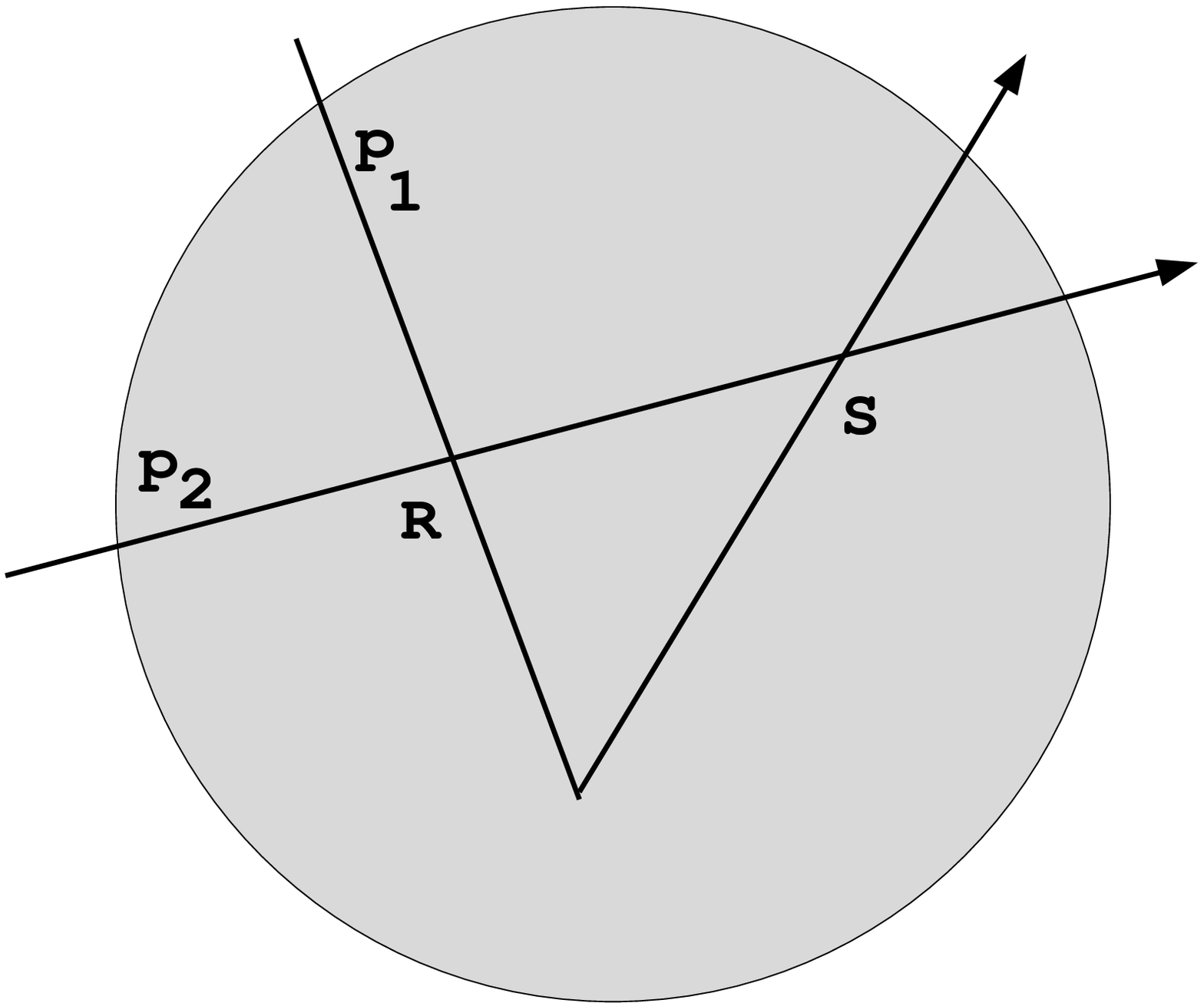}
\caption{ The V-intersection rule - the paths are shown in a shaded region of the plane, first 
separately and then together. One of the paths $p_1$ is crooked and intersects the straight path $p_2$ at 
two points $R$ and $S$ with opposite intersection numbers.} 

\label{Vintersection}
\end{figure}

\begin{figure}
\begin{center}
\includegraphics[width=20pc]{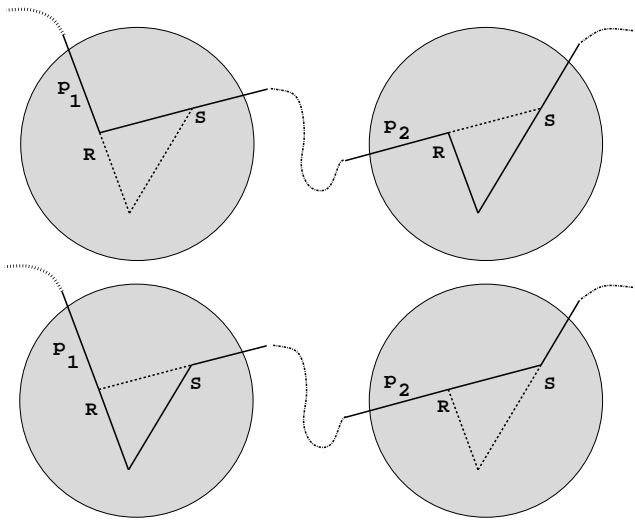}
\end{center}
\caption{ %(Zero-relative phase drawn on New Figures 1) 
The V-intersection rule - the relative phase between the 
two rerouting terms $p_1Rp_2$ and $p_1Sp_2$ resulting from Figure 
\ref{Vintersection} is zero, since the reroutings only differ in the position of one triangle, 
indicated by dotted lines, inside the shaded regions. The dashed lines outside the shaded regions 
represent the remainder of the paths $p_1$ and $p_2$.}
\label{zerorelphase}
\end{figure}

When part of the path $p_2$ is straight but part of the path $p_1$ is V-shaped, it can 
happen that two oppositely signed single intersections at points $R$ and $S$ occur, as 
shown in Figure \ref{Vintersection}.  Clearly the contributions from the
reroutings at $R$ and $S$ should cancel, since the path $p_2$ could be deformed
to avoid intersecting with $p_1$. However, when analyzing the corresponding
reroutings (see Figure \ref{zerorelphase}) it is seen that they have zero
relative area phase. Although the intersections have opposite sign, the two rerouting terms 
cannot cancel, since, in the refined Goldman bracket formula \rref{qgold}, the overall factor 
on the r.h.s. is $(q-1) + (q^{-1} - 1)\neq 0$. An identical argument shows that the cancellation also
fails to occur when the two paths are interchanged, i.e. $p_1$ is straight and
$p_2$ is V-shaped. To achieve this cancellation extra rules need to be applied.
There may be several ways of addressing this issue - we have chosen one possible approach.

The V-intersection rule is defined as follows. 

If the total intersection number $\epsilon(\overline{p_1},\overline{p_2})$ is
positive, an extra factor $q$ should be inserted into the term for the negative
rerouting, i.e. the overall factor for that rerouting, $(q^{-1} - 1)$, is to be
replaced by $q (q^{-1} - 1)  = - (q - 1)$. In this way this contribution will
cancel with that of the positive rerouting, since $(q-1) + q(q^{-1} - 1) = 0$.
This can be thought of as replacing $(q^{-1} - 1)$ with $- (q - 1)$, i.e.
changing the `incorrect' location of the negative intersection number in the exponent of $q$ to 
the `correct' location, i.e. as an overall sign multiplying the positive intersection number. 

Similarly, if the total intersection number $\epsilon(\overline{p_1},\overline{p_2})$ is 
negative, then an extra factor $q^{-1}$ should be inserted into the term for the 
`incorrect' positive rerouting, replacing $(q - 1)$ with  $- (q^{-1} - 1)$, or, again, 
changing the `incorrect' location of the intersection number in the exponent of $q$ to the `correct' 
location, and changing the overall sign. 

If the total intersection number $\epsilon(\overline{p_1},\overline{p_2})$ is
zero, then the contributions must cancel in pairs, and for each pair, either of
the $R$ and $S$ reroutings (but only one of them) should be adjusted with the
above rules.

\begin{figure}[h]
\begin{center}
\includegraphics[width=30pc]{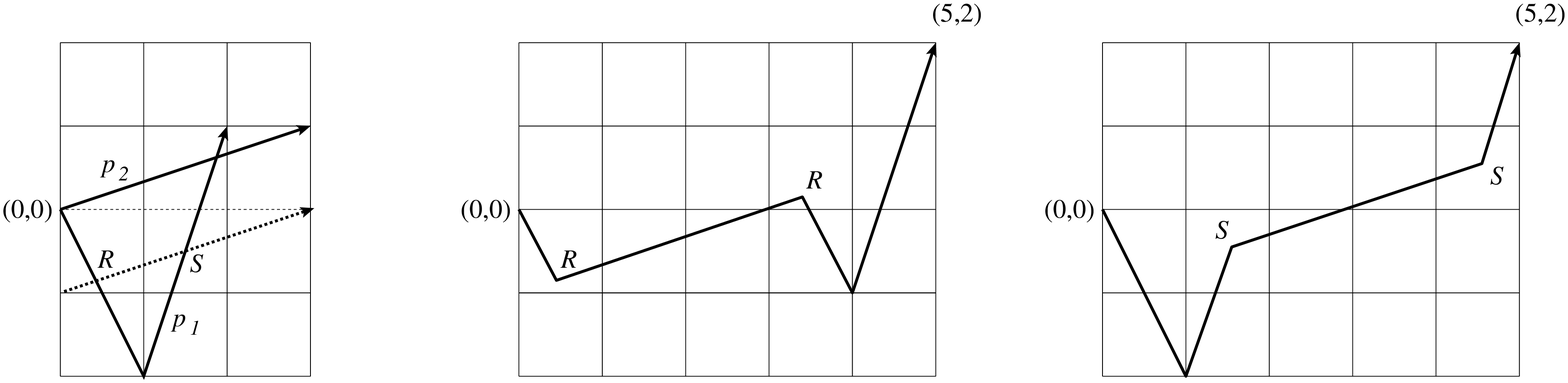}
\end{center}
\caption{\label{explicitV} The V-intersection rule - the paths $p_1$ and $p_2$ and the two reroutings 
$p_1Rp_2$ and $p_1Sp_2$.}
\end{figure}

As an example, take the $p_1$ of Figure \ref{zerorelphase} to be V-shaped,
starting at the origin $(0,0)$, passing through the point $(1,-2)$, and ending
at $(2,1)$ (i.e. $p_1$ is homotopic to the straight path $(2,1)$), whereas $p_2$
is straight, $p_2 = (3,1)$ (see the left figure in Figure \ref{explicitV}). Their total intersection number
$\epsilon(\overline{p_1},\overline{p_2})$ is $-1$, which would
correspond to a single intersection at the origin if the paths were straight.
Because $p_1$ is crooked, extra artificial intersections appear,
e.g. parallel translating $p_2$ to start at $(0,-1)$ we get intersections at 
$R$ (intersection number $+1$) and at $S$ (intersection number $-1$), giving rise to 
two rerouted paths which both end at the point $(5,2)$.  They are shown in Figure 
\ref{explicitV}. Because the overall intersection number is negative, the $R$ rerouting 
term should be adjusted with an extra factor $q^{-1}$.

Explicitly, applying the refined formula \rref{qgold} in this example gives 
$$
[\hat{T}(p_1), \hat{T}(p_2)] = q^{-1} (q-1) \hat{T} (p_1{R}p_2) + (q^{-1} -1) \hat{T} (p_1{S}p_2) + \cdots
$$
where the dots indicate the remaining positive rerouting terms and all the negative rerouting terms.  Indeed 
the first two terms
on the r.h.s. cancel as they should, due to the extra factor $q^{-1}$ in the first term coming from the 
V-intersection rule.

\vskip 10 pt

\subsection{Rule two - the regularization rule}\label{sub42}

\begin{figure}
\begin{center}
\includegraphics[width=8cm]{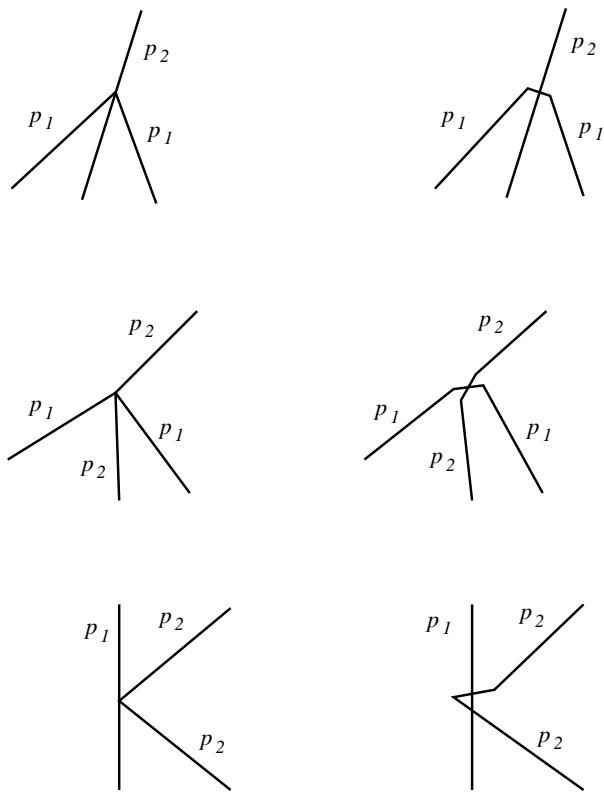}
\end{center}
\caption{\label{shiftreg} The regularization rule - the intersection point occurs where a crooked path 
(or paths) changes direction. The small line segments inserted in the figures on the right extend a 
distance $\epsilon$ to 
either side. }
\end{figure}

Intersections at points where one or both paths change direction precisely at those
points are ambiguous, since there is no longer a well-defined tangent direction at the intersection point 
for one or both paths - see the three examples on the
left in Figure \ref{shiftreg}. This may also occur when the
intersection point is the origin, and the direction of the path at its endpoint 
does not coincide with its direction at the starting point, i.e. the origin (an example is the 
intersection at the origin of the paths in Figure \ref{explicitV} of Section \ref{sub41}, another example 
will be discussed in Example 1, Section \ref{sub44}). The regularization rule consists 
in replacing the point where the path (or paths) changes direction with a
small line segment, extending a distance $\epsilon$ to either side of the point,
and adjusting the directions of the incoming and outgoing segments slightly in
accordance with this replacement - see the three examples of regularization on
the right in Figure \ref{shiftreg}. Note that the third example in Figure \ref{shiftreg} is unlike the 
other two in that the original, single intersection is replaced by two intersections, with opposite 
intersection number. An example of this type is the intersection point $U$ of Example 1, Section 
\ref{sub44} where we will show that the two rerouting terms that arise cancel due to the V-intersection rule.

The direction of the small line segment, or segments, should be chosen so that
the intersection with the other path is transversal. The effect of the insertion
of $\epsilon$ segments is to remove the ambiguity at the intersection points.
After analyzing the intersections and reroutings using the regularized paths,
the answer is given by the limit $\epsilon \rightarrow 0$.  

\begin{figure}
\begin{center}
\includegraphics[width=17pc]{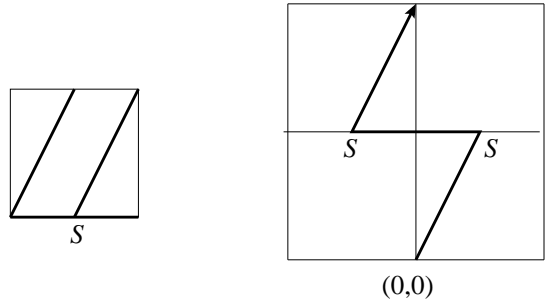}
\end{center}
\caption{\label{newregulariznexample1} The regularization rule - the path $p_1=(1,2)S(-1,0)$. }
\end{figure}

An example of this rule is as follows: take $p_1$ to be the crooked path
$p_1=(1,2)S(-1,0)$ (obtained by rerouting the straight path $(1,2)$ along the
straight path $ (-1,0)$ at the intersection point $S$ as shown on the r.h.s. of
Figure \ref{newregulariznexample1}), i.e. $p_1$ is homotopic to $(0,2)$, and
take $p_2$ to be the straight path $p_2= (3,2)$. The overall intersection number
is $\epsilon(\overline{p_1},p_2)=-6$, and corresponds to 6 integer starting points for parallel-translated 
copies of $p_2$
in the straight - straight pre - parallelogram, i.e. that formed by 
$\overline{p_1} = (0,2)$ and $\overline{p_2}=p_2=(3,2)$ (not shown), namely with vertices at
$-\overline{p_2}$, $-\overline{p_2} +\overline{p_1}$, $\overline{p_1}$ and the
origin $O$. 

\begin{figure}
\begin{center}
\includegraphics[width=15pc]{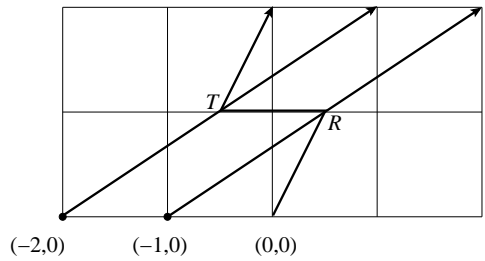}
\end{center}
\caption{\label{newregulariznexample2} The regularization rule - intersections between $p_1$ 
and $p_2=(3,2)$ at points $R$ and $T$ where $p_1$ changes direction. }
\end{figure}

In Figure \ref{newregulariznexample2} two of these integer starting points
are indicated by black dots, They are the points $(-1,0)$ and $(-2,0)$, and correspond 
to intersections at $R$ and $T$ respectively. At these points $R$ and $T$ the crooked path 
$p_1$ changes direction (see the first example of Figure \ref{shiftreg}) and also intersects 
the parallel copies of $p_2$ which start at $(-1,0)$ and $(-2,0)$. Therefore $p_1$ should be 
regularized at these intersections. At the remaining four intersection points $p_1$ does not 
change direction so no regularization is needed.

\begin{figure}
\begin{center}
\includegraphics[width=15pc]{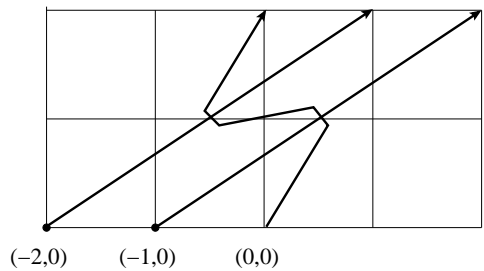}
\end{center}
\caption{\label{newregulariznexample3} The regularization rule - Figure \ref{newregulariznexample2} but with  
$p_1$ regularized. This gives transversal intersections. }
\end{figure}

\begin{figure}
\begin{center}
\includegraphics[width=25pc]{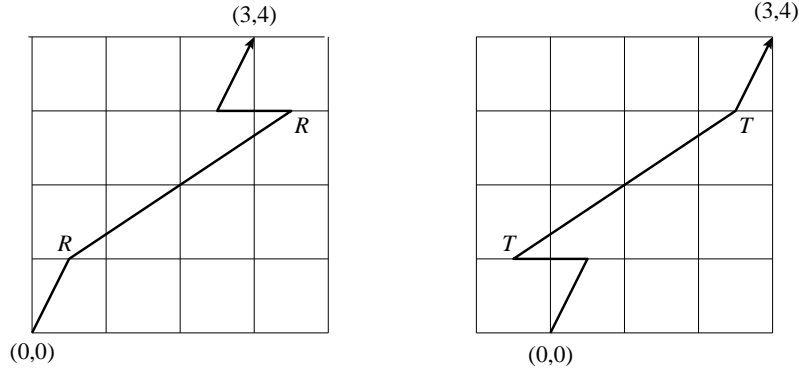}
\end{center}
\caption{\label{newregulariznexample4} The regularization rule - the reroutings $p_1 R p_2$ and $p_1 T p_2$. }
\end{figure}

Figure \ref{newregulariznexample3} shows the path $p_1$ regularized at $R$ and $T$, with small
segments of length $\epsilon$ inserted on either side of the intersections, to make them
transversal. Figure \ref{newregulariznexample4} displays the reroutings $p_1 R p_2$ and 
$p_1 T p_2$ in the limit $\epsilon \rightarrow 0$. 

\begin{figure}
\begin{center}
\includegraphics[width=10pc]{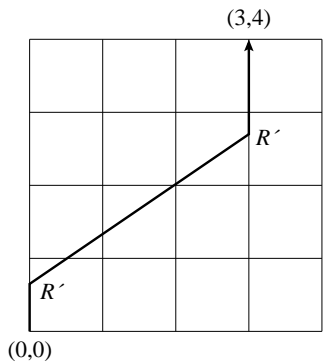} 
\end{center}
\caption{\label{newregulariznexample6} The regularization rule - the path $\overline{p_1}R'p_2$ i.e. 
rerouting at $R'$.}
\end{figure}

\begin{figure}
\begin{center}
\includegraphics[width=10pc]{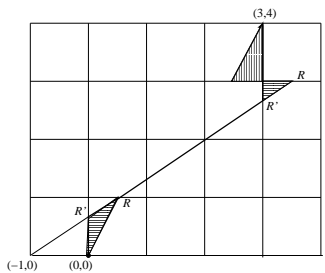}
\end{center}
\caption{\label{newregulariznexample5} The regularization rule - the relative area phase for $p_1Rp_2$ and 
$\overline{p_1}R'p_2$ is zero.}
\end{figure}

Combining these rerouting terms with the four terms from the other intersections, we find an 
expression for the (positive rerouting terms of the) commutator of  $\hat{T}(p_1)$ and
$\hat{T}(p_2)$ that is consistent with the commutator of
$\hat{T}(\overline{p_1})$ and $\hat{T}(\overline{p_2})=\hat{T}(p_2)$. 
We illustrate this statement in the case of the intersection point $R$. The commutators
should be equal, i.e.
\be
[\hat{T}(p_1), \hat{T}(p_2)] = [\hat{T}(\overline{p_1}), \hat{T}(p_2)]
\label{reg-exp-calc1}
\ee
since the relative area phase between $p_1$ and $\overline{p_1}=(0,2)$ is clearly zero
from Figure \ref{newregulariznexample1}. From the regularization rule as applied in 
Figure  \ref{newregulariznexample3}, the intersection number $\epsilon(p_1,p_2,R)$ 
is $-1$ which coincides with $\epsilon(\overline{p_1},p_2,R')$, where $R'$ is the 
intersection point corresponding to the integer starting point $(-1,0)$ in the 
pre - parallelogram. The rerouting $\overline{p_1}R'p_2$ is shown in Figure 
\ref{newregulariznexample6}.  We check that the positive rerouting terms in \rref{qgold}, 
corresponding to $R$ and $R'$ respectively, are in agreement with 
\rref{reg-exp-calc1}, i.e. equal. Indeed, as shown in Figure \ref{newregulariznexample5},
the relative area phase for these two reroutings is zero, since the sum of the horizontally 
shaded areas is clearly equal to the vertically shaded area. 

Therefore 
\be
(q^{\epsilon(p_1,p_2,{R})} - 1)\hat{T}(p_1{R}p_2)  =
(q^{\epsilon(\overline{p_1},p_2,{R'})} - 1)\hat{T}(\overline{p_1}{R'}p_2).
\label{reg-exp-calc2}
\ee

A similar reasoning applies to the intersection point $T$ and to the remaining four 
intersection points which need no regularization.

\subsection{Rule three - the missing rule}\label{sub43} 
\begin{figure}
\begin{center}
\includegraphics[width=9cm]{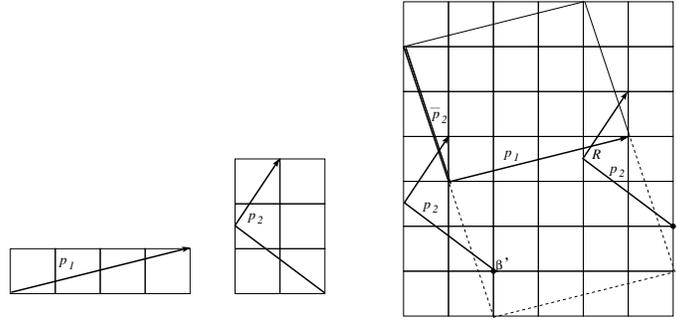}
\end{center}
\caption{%(Missing rule on New Figures 2) 
The missing rule - illustration. }
\label{missingrule}
\end{figure}

The final rule, called the missing rule, occurs if we parallel-translate $p_2$
to start at an integer point $\beta^{\prime}$ inside the pre - parallelogram for the
straightened paths $\overline{p_1}$ and $\overline{p_2}$, and then find that it
does not intersect (i.e. misses) the first path $p_1$. An example is shown in  
Figure \ref{missingrule}, where $p_1$ is straight and $p_2$ is crooked. Note
that although $p_2$ is V--shaped the missing rule rule should not be confused
with the V--intersection rule, which concerns the cancelling of reroutings from
pairs of oppositely-signed intersections. Translating $p_2$ to start at the
integer point $\beta'$, lying inside the pre - parallelogram for the straightened
paths $\overline{p_1} = p_1$ and $\overline{p_2}$ (dotted lines), it does not
intersect $p_1$. Hence the usual analysis of intersections and reroutings does
not apply. In such circumstances we apply a `missing rule', i.e. add an integer
multiple of $\overline{p_1}$  to $\beta'$ to get a new integer starting
point $\beta$, outside the pre - parallelogram, such that the parallel-translated 
copy of $p_2$ starting at $\beta$ does intersect $p_1$. In Figure \ref{missingrule} 
this is achieved by choosing $\beta = \beta' + p_1$. The resulting intersection 
point is denoted by $R$. 

In such cases the term corresponding to the rerouting  of $p_1$ along $p_2$  at
the intersection point $R$ is given an adjustment factor:
\be
q^{\det(\beta^{\prime} - \beta, \overline{p_2})}
% - \det(\beta, \overline{p_2})}
\label{miss}
\ee
to compensate for the fact that its area phase relative to the ``normal'' rerouted
paths has been altered, where ``normal'' refers to those rerouted paths which arise 
from a parallel copy of $p_2$ with starting point inside the straight - straight pre - parallelogram.

\begin{figure}
\begin{center}
\includegraphics[width=5cm]{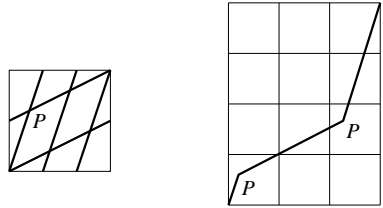}
\end{center}
\caption{The missing rule - the fundamental reduction of straight paths $(1,3)$ and $(2,1)$ 
and the rerouting $p_2=(1,3)P(2,1)$. }
\label{newmissingexample1}
\end{figure}

For example, consider the two paths $p_1=(-1,0)$ (straight) and
$p_2=(1,3)P(2,1)$, a crooked path obtained by rerouting $(1,3)$ along $(2,1)$ at
the intersection point $P$ - see 
Figure \ref{newmissingexample1}. 

\begin{figure}
\begin{center}
\includegraphics[width=4cm]{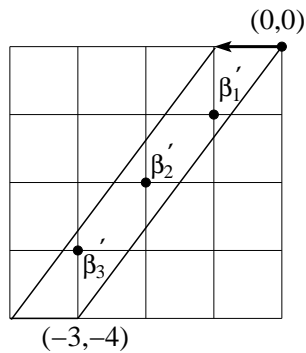}
\end{center}
\caption{The missing rule - the straight - straight pre - parallelogram for $p_1=(-1,0)$ and $\overline{p_2}=(3,4)$.}
\label{newmissingexample2}
\end{figure}

The straightened path $\overline{p_2}$ is $(3,4)$ and the total intersection
number between $p_1$ and $p_2$ is $\det(p_1, \overline{p_2}) =-4$. Figure 
\ref{newmissingexample2} shows the straight - straight pre - parallelogram (namely that with 
clockwise vertices $-\overline{p_2}$, $-\overline{p_2} +\overline{p_1}$, $\overline{p_1}$ 
and the origin $O=(0,0)$). There are four integer starting points, namely the origin 
$O=(0,0)=\beta^{\prime}_O$, $\beta^{\prime}_1=(-1,-1)$, $\beta^{\prime}_2=(-2,-2)$ and $\beta^{\prime}_3=(-3,-3)$. 

\begin{figure}
\begin{center}
\includegraphics[width=3cm]{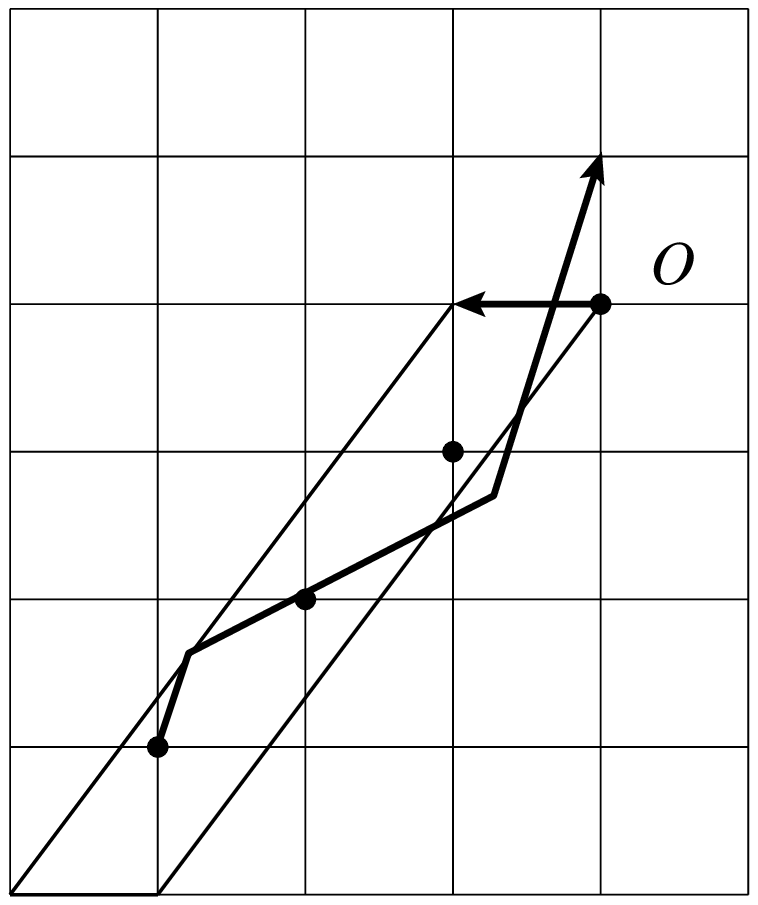}
\hspace{0.5cm}
\includegraphics[width=3cm]{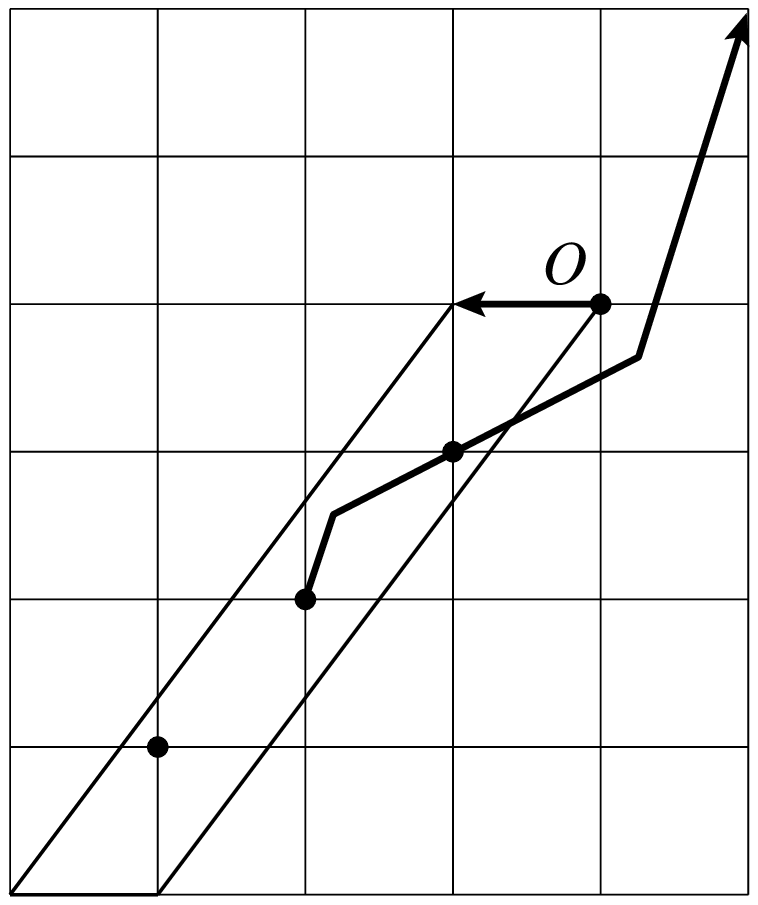}
\hspace{0.5cm}
\includegraphics[width=3cm]{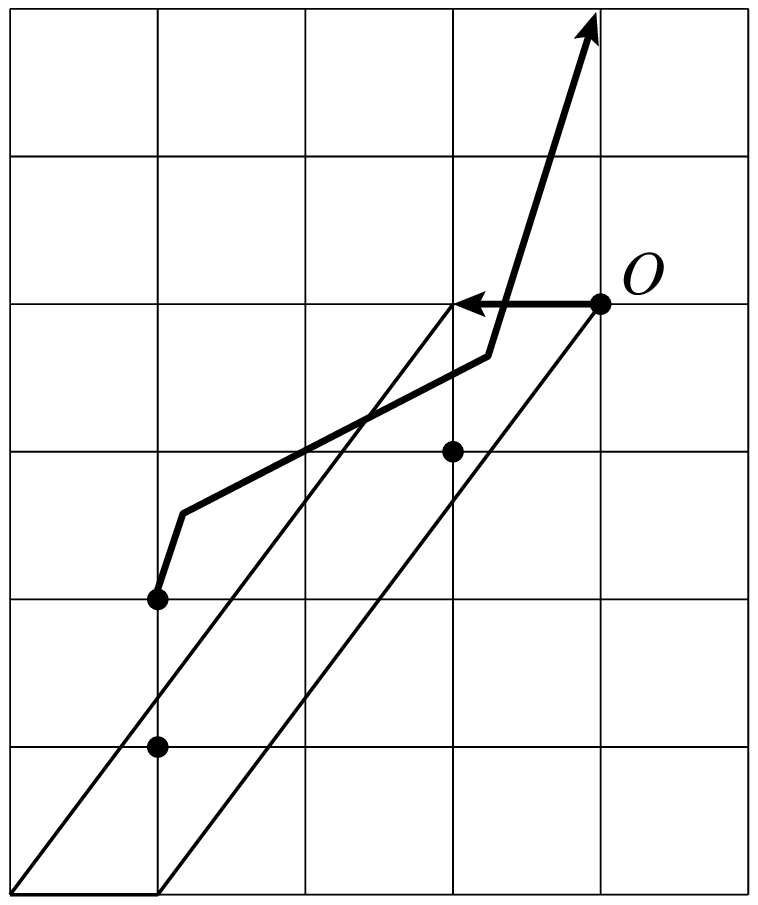}
\end{center}
\caption{The missing rule - Figure \ref{newmissingexample2} with parallel-translated copies of $p_2$ 
starting at $(-3,-3)$, $(-2,-2)$ and $(-3,-2)$ respectively.}
\label{newmissingexample3}
\end{figure}

Parallel translating $p_2$ to start at these integer points we find that in
three cases, namely starting points $O=\beta^{\prime}_O$, $\beta^{\prime}_1$ and $\beta^{\prime}_3$, we 
get an
intersection with $p_1$. See for instance the figure on the left of Figure
\ref{newmissingexample3} for the starting point $\beta^{\prime}_3$ - the other two
starting points have the same property. However in one case, namely starting
point $\beta^{\prime}_2=(-2,-2)$, there is no intersection between the parallel-translated copy
of $p_2$ starting at $\beta^{\prime}_2$ and $p_1$ (see the central 
figure of Figure
\ref{newmissingexample3}).  Applying the missing rule we take the new starting
point to be $\beta_2 = (-3,-2) = p_1 + \beta^{\prime}_2$, as shown in the r.h.s. figure  of Figure
\ref{newmissingexample3}, and compensate by multiplying the corresponding
rerouting by the factor \rref{miss} where $\beta^{\prime}_2 - \beta_2 = - p_1 =(1,0)$ and 
$\overline{p_2} = (3,4)$, i.e. $\det(\beta^{\prime}_2 - \beta_2, \overline{p_2}) = 4$, and the 
compensating factor is $q^4$.

We now illustrate how the missing rule gives a result consistent with the
straight - straight result for this example. The area phase for $p_2$ relative to $\overline{p_2}$ is
$$
S(p_2, \overline{p_2}) = S((p_2, (3,4)) = 3/2,
$$
as may be seen by using as an intermediary the polygonal path $p_3$ going in straight segments 
from $(0,0)$ to $(1,3)$ to $(3,4)$:
$$
S(p_2, p_3) = 4, \quad \quad S(p_3, (3,4)) = -(2+\frac{3}{2} -1) = -5/2,
$$
where we used Pick's formula \rref{pick} in the second equation. Then we can use the unrefined formula 
\rref{qgb1} to calculate the commutator:
\begin{eqnarray}
[\hat{T}(p_1),\hat{T}(p_2)] & = & q^{3/2}[\hat{T}(-1,0),\hat{T}(3,4)] \nonumber\\
&  = & q^{3/2} (q^{-2}- q^{2})(\hat{T}(2,4) - \hat{T}(4,4)).
\label{t12}
\end{eqnarray}

We show how one can recover the positive rerouting term in \rref{t12} from the refined formula \rref{qgold} 
together with the missing rule: 
\begin{eqnarray}
\lefteqn{\sum_{ {I} \in p_1 \sharp p_2} (q^{\epsilon(p_1,p_2,{I})} - 1) \hat{T}(p_1{I}p_2) } \nonumber\\
& = & (q^{-1} -1)\, \hat{T}(p_1{O}p_2) \left( 1 + q^{\det(\beta^{\prime}_1, \overline{p_2})}
+q^4 . \,  q^{\det(\beta_2, \overline{p_2})} + q^{\det(\beta^{\prime}_3, \overline{p_2})} \right) \nonumber\\
& = & (q^{-1} -1) \, \hat{T}(p_1{O}p_2) \, \left( 1+ q^{-1} + q^{-2} + q^{-3}  \right) \nonumber\\
& = & (q^{-4} -1)\, \hat{T}(p_1{O}p_2) \nonumber\\
&  = & q^{3/2} \,q^2 \, (q^{-4} -1)\, \hat{T}(2,4) = q^{3/2} \, (q^{-2}- q^{2})\, \hat{T}(2,4).
\label{t13}
\end{eqnarray}

In the first equality of \rref{t13} we have used the signed area equation \rref{s3} to relate three 
rerouting terms to the rerouting term at the origin $\hat{T}(p_1{O}p_2)$. The adjustment factor from 
the missing rule appears as a multiplier in the 3rd term. The passage from  $\hat{T}(p_1{O}p_2)$ to 
$\hat{T}(2,4)$ (the last line of \rref{t13}) comes from 
$$
\hat{T}(p_1{O}p_2) = q^{3/2}\, \hat{T}(p_1{O}\overline{p_2})= q^{3/2}\, q^2 \, \hat{T}(2,4)
$$
where $p_1{O}\overline{p_2}$ is the polygonal path rerouted at the origin, i.e. straight 
segments from $(0,0)$ to $(3,4)$ to $(2,4)$, or $\overline{p_2}=(3,4)$ followed by $p_1=(-1,0)$.

\vskip 10 pt
\subsection{Two examples of combinations of rules}\label{sub44}

{\bf Example 1.} 

\begin{figure}[hbpt]
\centering
\includegraphics[height=3cm]{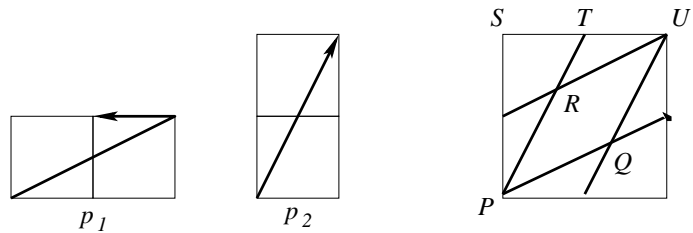}
\caption{Example 1: the two paths and the reduced paths in a fundamental domain.} 
\label{shiftexample1}
\end{figure}

This example requires both the V--intersection rule and the
regularization rule. Take $p_1$ to be the crooked polygonal line connecting
$(0,0)$, $(2,1)$ and $(1,1)$ in that order, and $p_2$ to be the straight path
$p_2=(1,2)$. Their total intersection number is $+1$, although from Figure
\ref{shiftexample1} (the two paths and the reduced paths in a fundamental
domain) it appears that there are 6 intersection points. Clearly 5 of these are
{\it artificial}, since $p_1$ is crooked, and their contributions must
cancel. This is possible, since after regularization there are six artificial 
intersection points (see the third example in Figure 
\ref{shiftreg}), and they cancel in pairs. 

The pair $R$ and $T$ will cancel due to the V--intersection rule,
without regularization, and the pair $Q$ and $S$ will cancel, again due to the
V--intersection rule, but after regularization. The intersection point $U$ is a
case where the regularization rule must be applied as in the third example of Figure 
\ref{shiftreg}. After regularization, there are actually two intersection
points, denoted $U_1$ and $U_2$, between the regularized path $p_1$ and $p_2$,
and the corresponding rerouting terms cancel when applying the V-intersection
rule,  The remaining intersection at $P$ must be regularized. We now proceed to
analyse these statements in more detail.

\begin{figure}[hbpt]
\centering
\includegraphics[height=3cm]{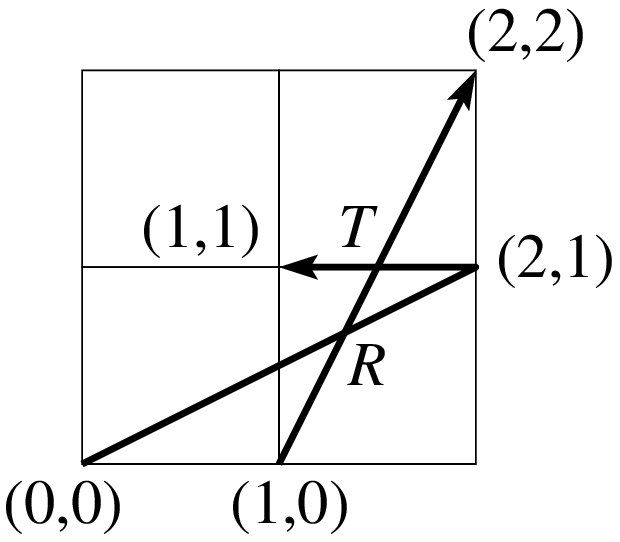}
\hspace{1cm}
\includegraphics[height=4cm]{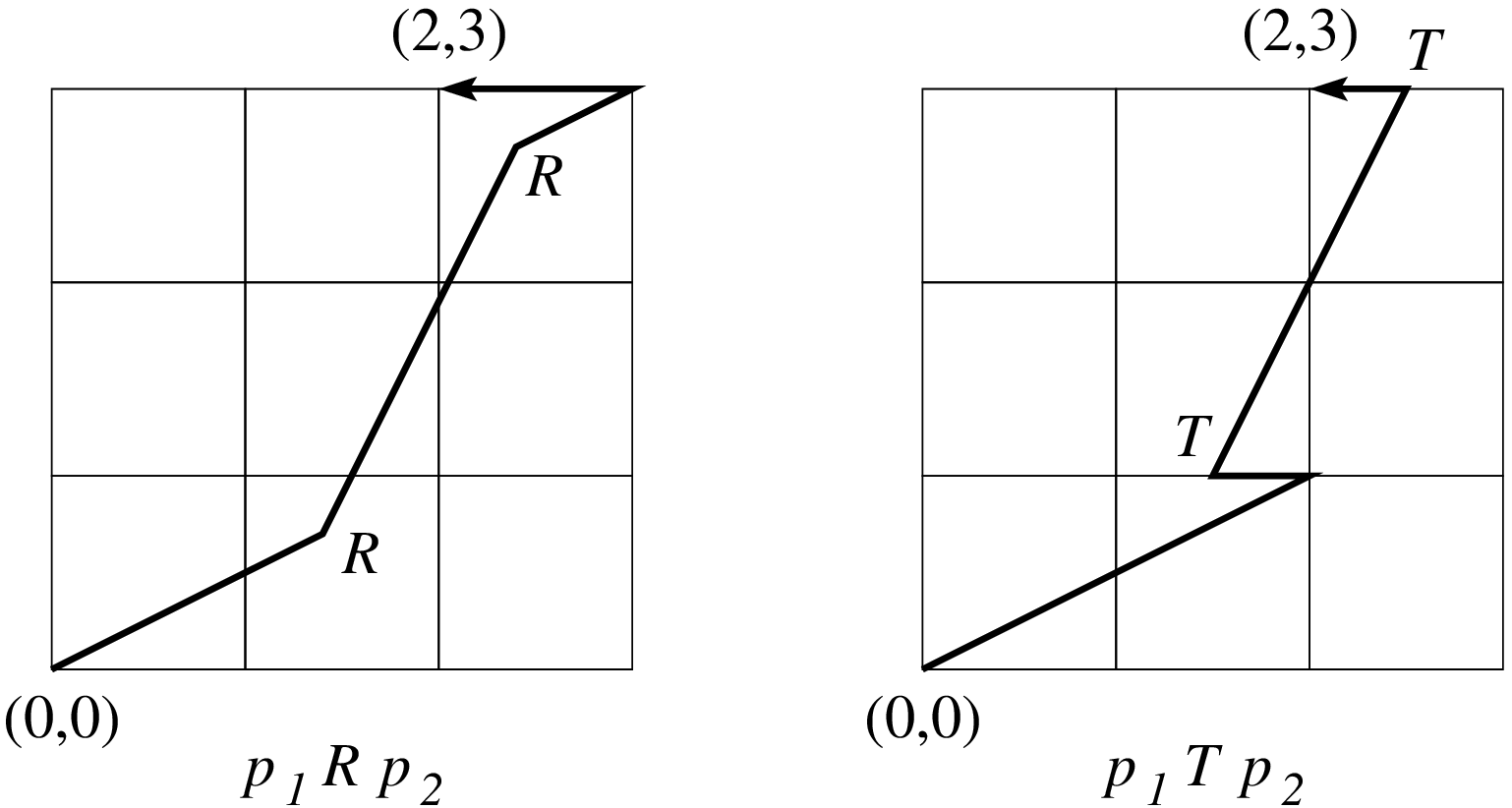}
\caption{Example 1: the V-intersection rule 1 is applied.} 
\label{rule1example}
\end{figure}

The pair $R$ and $T$ - the intersection points $R$ (intersection number
$+1$) and $T$ (intersection number $-1$) form a pair since one can
consider them as arising artificially from shifting $p_2$ to start at the point
$(1,0)$, and end at the point $(2,2)$, as shown on the left in Figure
\ref{rule1example}. There we have also displayed the corresponding reroutings
$p_1Rp_2$ and $p_1Tp_2$, and clearly they have relative signed area, namely
zero, with respect to each other, or equivalently, they have the same relative
signed area, namely $2$, with respect to the straight reference path with
endpoint at $(2,3)$. 

The V-intersection rule must be applied to $\hat{T}(p_1Tp_2)$, giving an extra
factor $q$, since it has the wrong sign with respect to the total intersection
number. The two terms corresponding to $R$ and $T$ in the refined bracket
$[\hat{T}(p_1),\hat{T}(p_2)]$ then indeed cancel, 
$$
(q-1)\hat{T}(p_1Rp_2) + q(q^{-1}-1)\hat{T}(p_1Tp_2) = ((q-1) + (1-q))q^2 \hat{T}(2,3)=0.
$$

\begin{figure}
\begin{center}
\includegraphics[width=8cm]{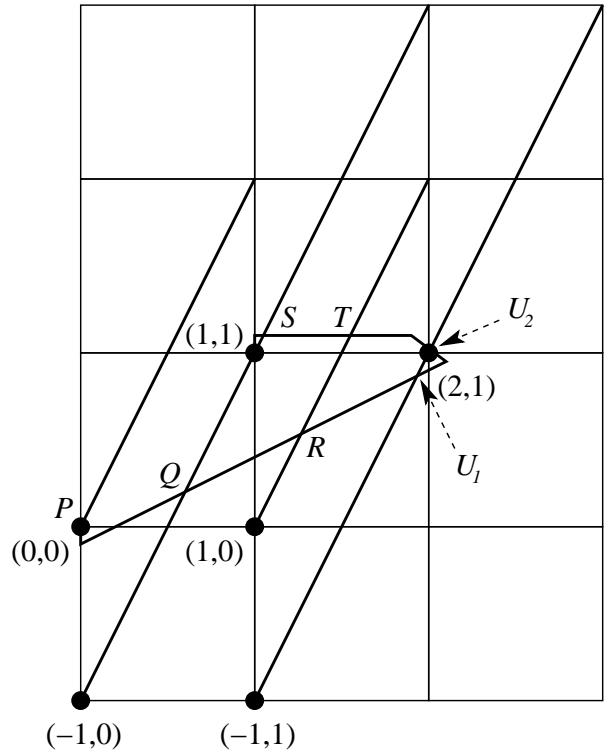} 
\end{center}
\caption{\label{shiftexample2} Example 1: the regularization rule is applied to the path $p_1$ 
(crooked) at the points $(0,0)$ and $(2,1)$, and 
the parallel-translated copies of $p_2$ (straight), with starting points indicated by black dots, 
give rise to intersections.}
\end{figure}

The $P$ intersection at the origin is an intersection point, but $p_1$ changes direction
there since the directions at its starting and endpoints do not coincide: the
tangent vector at its starting point is $(2,1)$ and at its endpoint the tangent
vector is $(-1,0)$. It should be regularized. We therefore insert a small line
segment in the path $p_1$ at the origin, and look at the intersections - see
Figure \ref{shiftexample2}. The four dots are integer starting points for the
parallel-translated copies of $p_2$. The origin is the starting point for the intersection at $P$ (and 
is also $P$ itself), $(1,0)$ is the starting point for the intersections at $R$ and $T$, $(0,-1)$ is 
the starting point for the intersection at $Q$, and $(1,1)$ is the starting point for the intersection at $S$.

\begin{figure}
\begin{center}
\includegraphics[height=4cm]{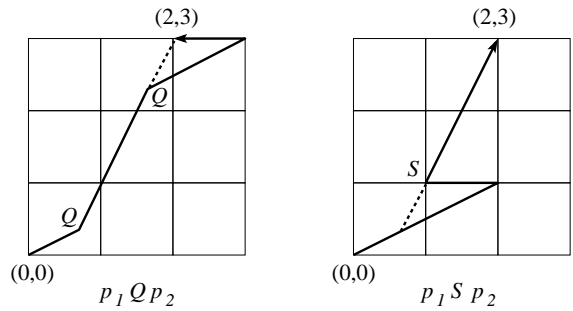}
\end{center}
\caption{\label{shiftexample3} Example 1: the reroutings $p_1Qp_2$ and $p_1Sp_2$ have relative area 
phase 
zero, since they only differ in the position of the triangle indicated by the dotted lines. }
\end{figure}

The $Q$ and $S$ intersections also cancel, in the limit $\epsilon \rightarrow 0$. This should be 
clear 
from Figure \ref{shiftexample3}, where again we invoke the V-intersection rule.

\begin{figure}
\begin{center}
\includegraphics[height=4cm]{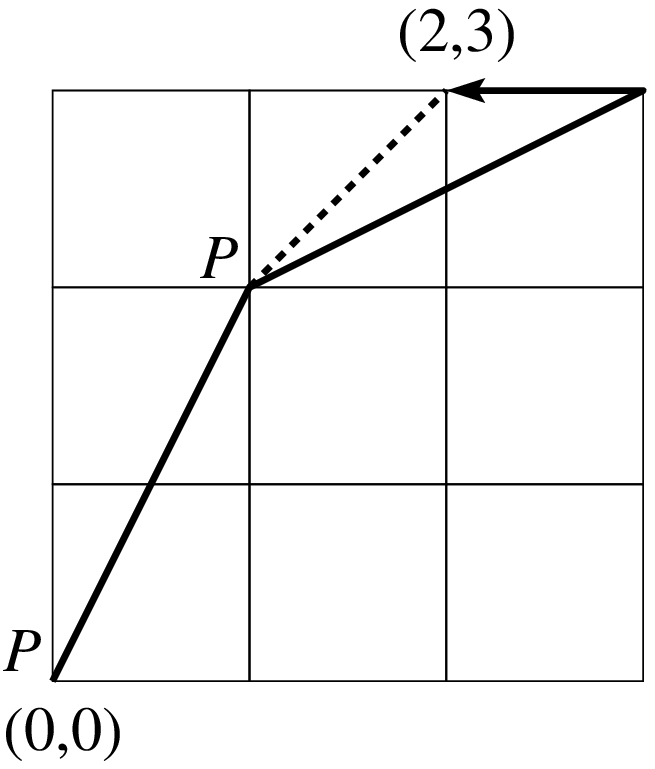}
\end{center}
\caption{\label{shiftexample4} Example 1: the rerouting $p_1Pp_2$ shown as a solid line and the 
rerouting $\overline{p_1} P p_2 $, which goes from the origin to $(1,2)$, then follows the dashed 
line to $(2,3)$.}
\end{figure}

The intersection at $U$ is more subtle since the path $p_1$ also changes direction there (see Figure 
\ref{shiftexample1}), so we should apply the regularization rule. It is shown in Figure
\ref{shiftexample2} that this leads to two intersections, denoted $U_1$ and $U_2$,
with opposite intersection numbers. When applying the V-intersection rule, the
corresponding reroutings (not shown) cancel, since they are equal apart from the
intersection number factor and the adjustment factor which comes from applying
the V-intersection rule.

Thus the refined bracket $[\hat{T}(p_1),\hat{T}(p_2)]$ gives just one remaining
term (for the positive rerouting terms) namely $\hat{T}(p_1Pp_2)$, shown in 
Figure \ref{shiftexample4}. This is correct, since the relative signed area
compared to the rerouted path which would have arisen if we had used
$\overline{p_1}=(1,1)$ and $p_2$, also shown in  Figure \ref{shiftexample4}, is
exactly the same as the relative signed area relating $p_1$ and
the straightened $\overline{p_1}$ (namely $q^{1/2}$ in both cases). In summary, the refined bracket 
calculation for this example gives:
$$
[\hat{T}(p_1),\hat{T}(p_2)] = (q-1)\hat{T}(p_1 P p_2)= q^{1/2}(q-1)\hat{T}(\overline{p_1} P p_2)
$$
after the cancelling of six terms, which is consistent with the straight path result:
$$
[\hat{T}(\overline{p_1}),\hat{T}(p_2)] = (q-1)\hat{T}(\overline{p_1} P p_2)
$$
since $\hat{T}(\overline{p_1})= q^{1/2} \hat{T}(p_1)$.

\noindent {\bf Example 2.} 

This example requires both the regularization rule and the missing rule  Take $p_1$ to be $p_1 = (2,1)$, 
i.e. straight, and $p_2$ the crooked polygonal path from $(0,0)$ to  $(-2,1)$ to $(-1,2)$. Their 
total intersection number is $+5$.

\begin{figure}
\begin{center}
\includegraphics[width=12cm]{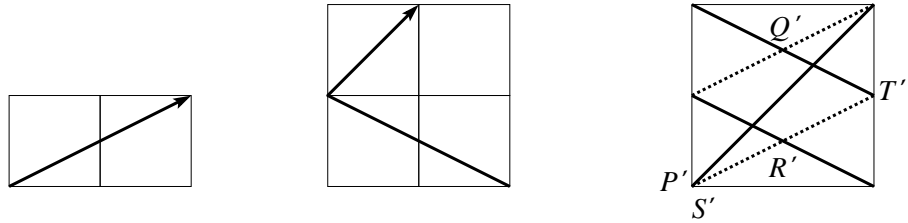}
\end{center}
\caption{\label{combined-exampleIc} Example 2: the paths $p_1$ and $p_2$, 
and their intersections ($p_1$ dotted, $p_2$ solid) in a fundamental domain.}
\end{figure}

\begin{figure}
\begin{center}
\includegraphics[width=8cm]{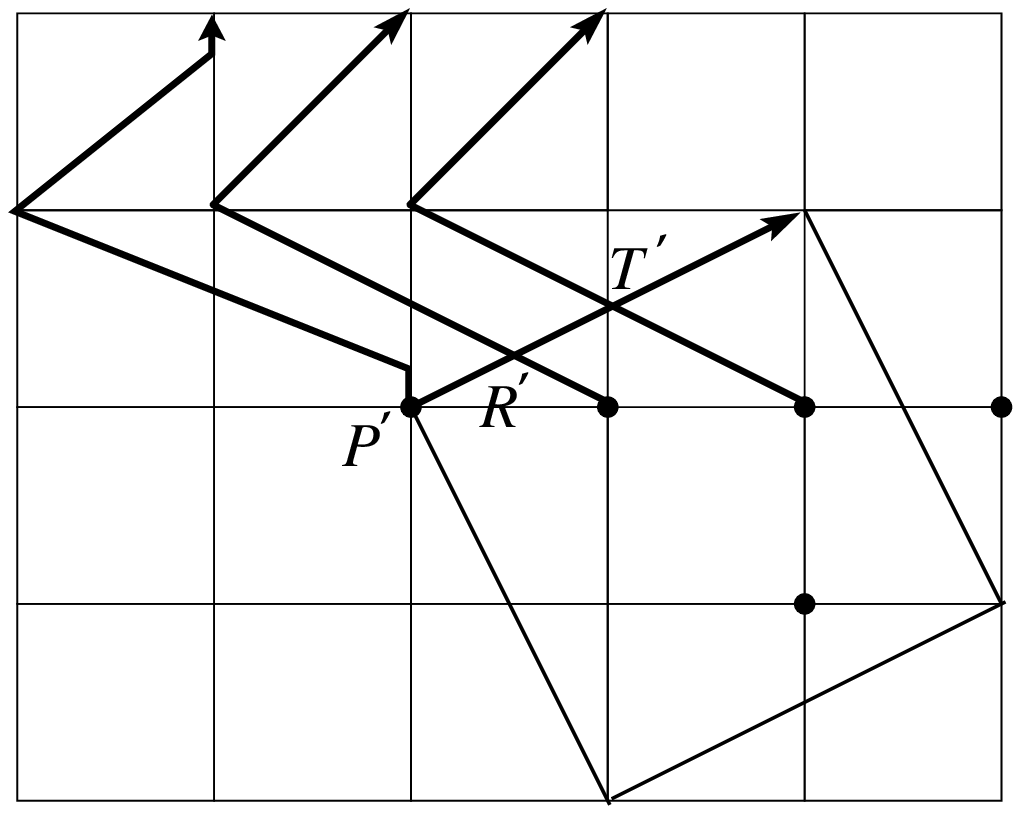}
\end{center}
\caption{\label{combined-exampleIe} Example 2: The intersections $P'$, $R'$ and $T'$ between $p_1$ and $p_2$ 
and the corresponding parallel-translated copies of ${p_2}$.}
\end{figure}

\begin{figure}
\begin{center}
\includegraphics[width=8cm]{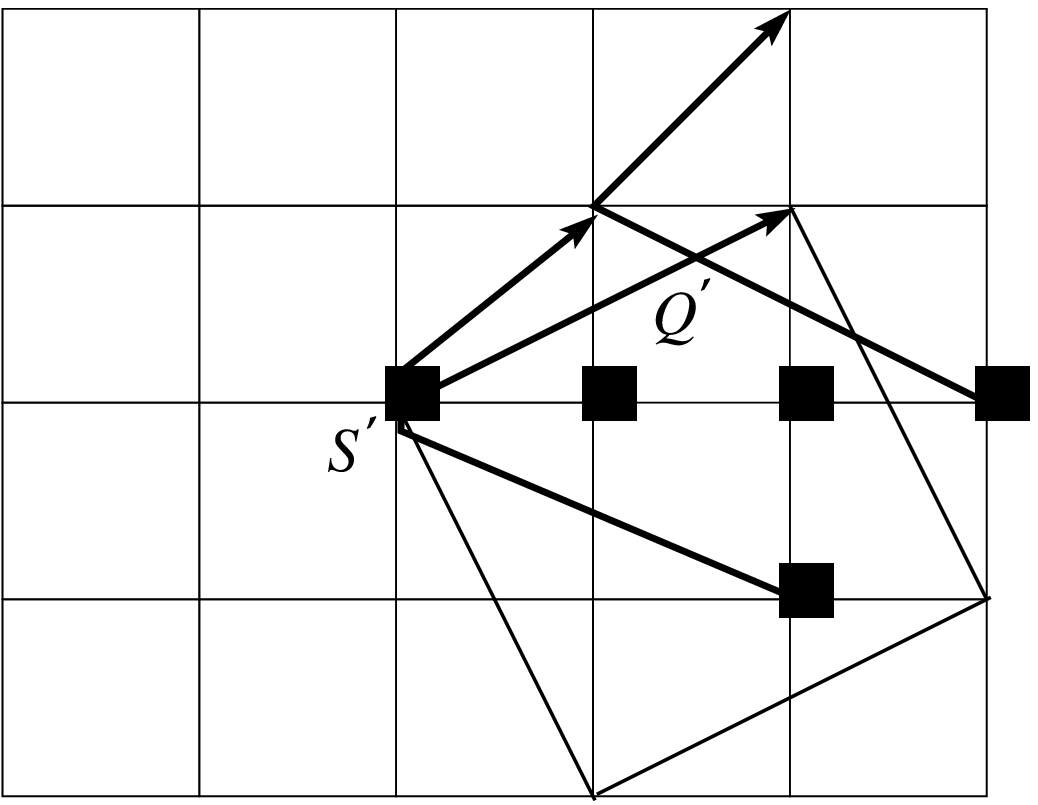}
\end{center}
\caption{\label{combined-exampleId} Example 2: The intersections $S'$ and $Q'$ between $p_1$ and 
$p_2$ and the corresponding parallel-tranlated copies of ${p_2}$.}
\end{figure}

Figure \ref{combined-exampleIc} shows the paths $p_1$ and ${p_2}$ (crooked) and their $5$ 
intersection points, $P'$, $Q'$,$R'$, $S'$, and $T'$, found by reducing to a fundamental domain. The 
intersection points $P'$ and $S'$ are both indicated at the bottom left-hand corner, since they 
both occur at the starting point of $p_1$ (but at two different points of $p_2$ - see 
Figures \ref{combined-exampleIe} and \ref{combined-exampleId} where the parallel transported copies 
of $p_2$ are displayed).

\begin{figure}
\begin{center}
\includegraphics[width=3cm]{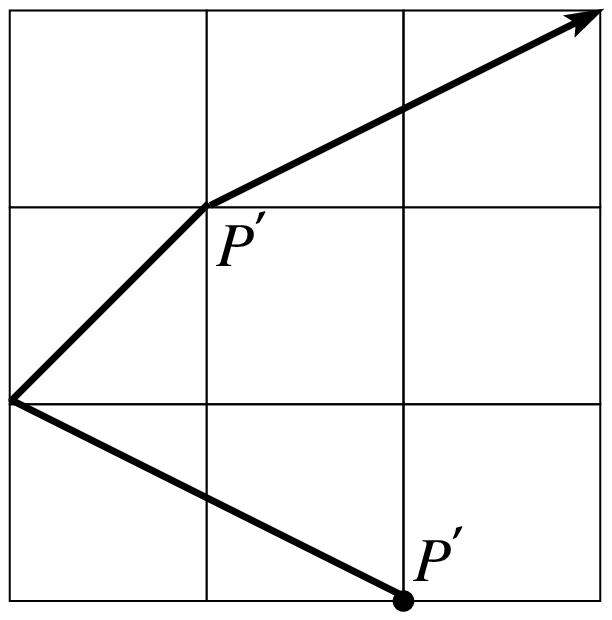}
\hspace{1cm}
\includegraphics[width=3cm]{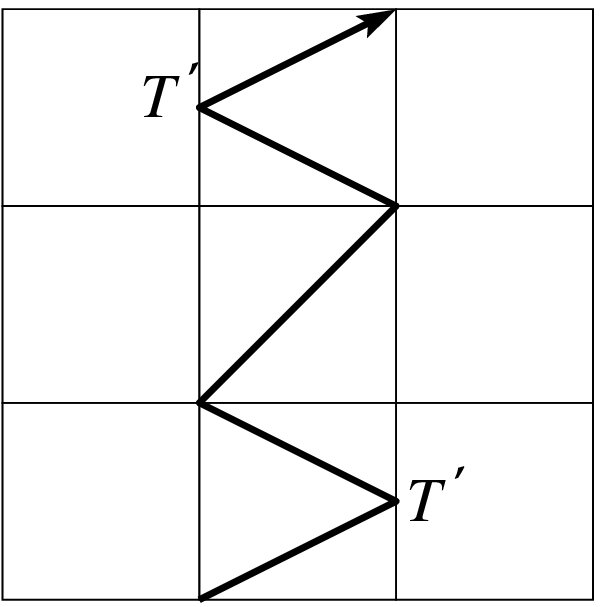}
\hspace{1cm}
\includegraphics[width=3cm]{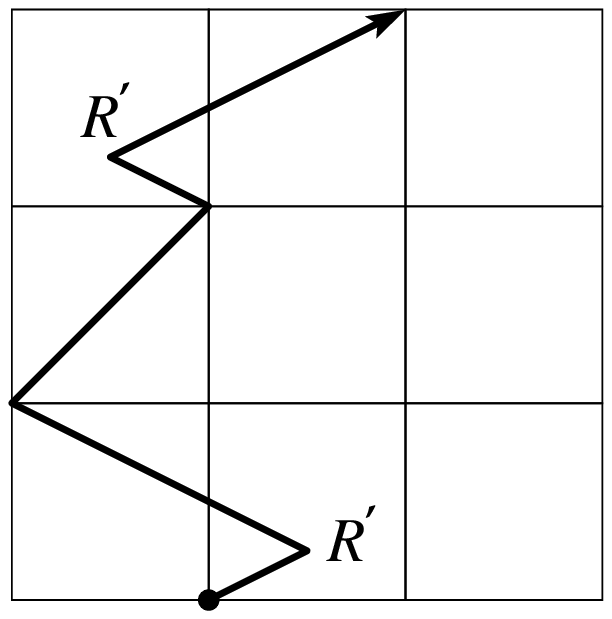}
\hspace{1cm}
\newline
\vspace{0.2cm}
\newline
\includegraphics[width=3cm]{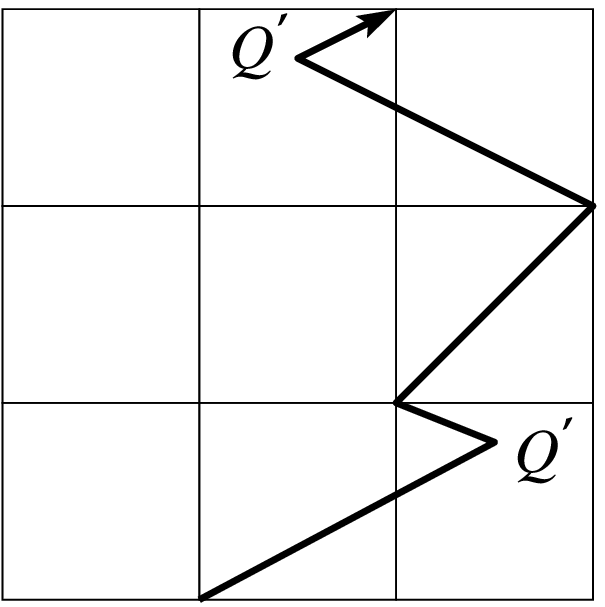}
\hspace{1cm}
\includegraphics[width=3cm]{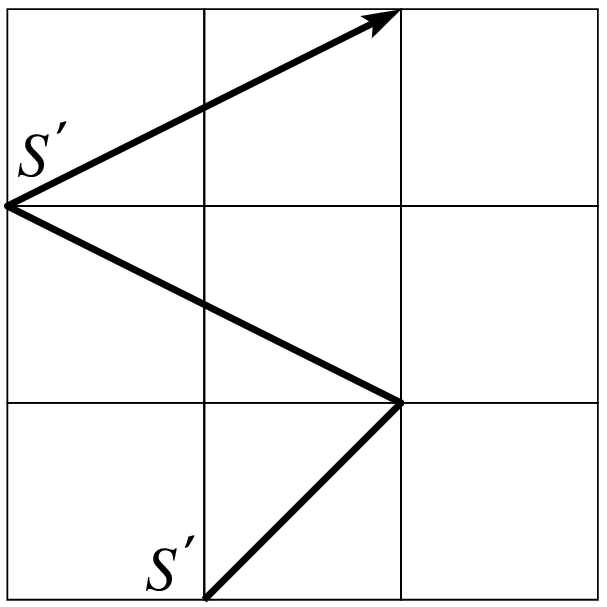}
\end{center}
\caption{\label{5reroutings} Example 2: the five positive reroutings arising from the intersections 
between $p_1$ and $p_2$.}
\end{figure}

Figure \ref{5reroutings} shows the five positive reroutings which arise from these intersections.

\begin{figure}
\begin{center}
\includegraphics[width=6cm]{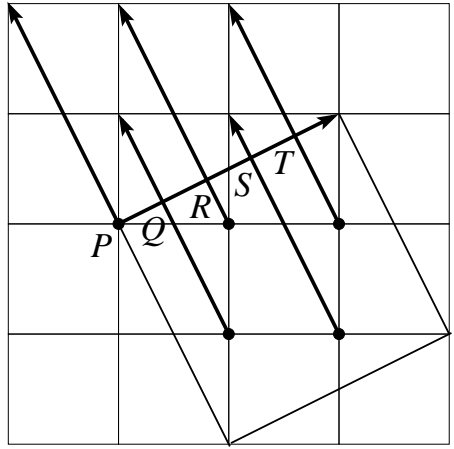}
\end{center}
\caption{\label{combined-exampleIb} Example 2: the pre - parallelogram for $p_1$ and $\overline{p_2}$.}
\end{figure}

Figure \ref{combined-exampleIb} shows the pre - parallelogram for $p_1$ and the
straightened path $\overline{p_2}= (-1,2)$. The $5$ intersection points, $P$, $Q$, $R$, $S$, and 
$T$, correspond to 5 integer starting points for parallel translated copies of $\overline{p_2}$ 
(parallel transported to start at integer points inside the pre - parallelogram.) 

We now analyse the intersections between $p_1$ and ${p_2}$ using
parallel - translated copies of ${p_2}$ starting at different integer points.

For three of the intersections, namely $P'$, $R'$ and $T'$, the integer
starting points coincide with starting points in the straight - straight
pre - parallelogram of Figure \ref{combined-exampleIb}, and $R'$ and $T'$ are transversal
intersection points. However $p_2$ has a different direction departing from and arrriving at the 
intersection point $P'$, so there the regularization rule needs to be applied - see Figure
\ref {combined-exampleIe}. 

For the intersection $S'$, the integer starting point also coincides with a starting point in the 
straight - straight pre - parallelogram of Figure \ref{combined-exampleIb}, but the intersection with $p_1$ is 
no longer transversal. This is because $S'$ is at the vertex of $p_2$, i.e. where $p_2$ changes direction. 
Therefore the tangent direction of $p_2$, and also the intersection number, are not well-defined there 
(see Figure \ref{combined-exampleId}). Again the regularization rule should be applied, as is done in that Figure. 

The intersection point $Q'$ is an instance where we must apply the missing rule, since 
when $p_2$ is parallel-translated to the fifth integer point in the straight - straight
pre - parallelogram of Figure \ref{combined-exampleIb}, it fails to intersect
$p_1$.

Figure \ref{combined-exampleId} shows the situation for the $S'$ and $Q'$ intersections, displaying
the regularization of the copy of $p_2$ for $S'$ and the integer starting point
outside the straight - straight pre - parallelogram for $Q'$. The adjustment factor that must be applied
to the corresponding rerouting term because of the missing rule (the $Q'$ intersection) is given by \rref{miss} with 
$\beta^{\prime} - \beta = - p_1 =(-2,1)$ and 
$\overline{p_2} = (-1,2)$, i.e. $\det(\beta^{\prime} - \beta, \overline{p_2}) = -5$, i.e. the 
compensating factor is $q^{-5}$.

Finally we show part of the calculation which confirms that the application of these rules is consistent with  
the straight - straight result for this example. 
The area phase for $p_2$ relative to $\overline{p_2}$ is
$$
S(p_2, \overline{p_2}) = S((p_2, (-1,2)) = -(1+3/2 -1) = -3/2
$$
where we used Pick's formula \rref{pick} in the second equality. Then we use the unrefined formula \rref{qgb1} 
to calculate the commutator:
\begin{eqnarray}
[\hat{T}(p_1),\hat{T}(p_2)] & = & q^{-3/2}[\hat{T}(2,1),\hat{T}(-1,2)] \nonumber\\
&  = & q^{-3/2} (q^{5/2}- q^{-5/2})(\hat{T}(1,3) - \hat{T}(3,-1)).
\label{ex2}
\end{eqnarray}

In \rref{ex2} the positive rerouting term i.e. $(q-q^{-4}) \hat{T}(1,3)$ can be recovered from the refined 
formula \rref{qgold}, after applying both the regularization rule and missing rules: 
\begin{eqnarray}
\lefteqn{\sum_{ {I} \in p_1 \sharp p_2} (q^{\epsilon(p_1,p_2,{I})} - 1) \hat{T}(p_1{I}p_2) } \nonumber\\
& = & (q -1)\, \hat{T}(p_1{O}p_2) 
\left( 
1 + q^{\det(\beta_{R'}, \overline{p_2})} + q^{\det(\beta_{T'}, \overline{p_2})}  \right. \nonumber\\
&  & \qquad \qquad \qquad \qquad \qquad \left.
+ q^{\det(\beta_{S'}, \overline{p_2})} + q^{-5}.q^{\det(\beta_{Q'}, \overline{p_2})}
\right) \nonumber \\
& = & (q -1) \, \hat{T}(p_1{O}p_2) \, \left( 1+ q^{2} + q^{4} + q^{3} + q^{-5}¬q^{6}\right) \nonumber\\
& = & (q^{5} -1)\, \hat{T}(p_1{O}p_2) \nonumber\\
& = & (q^{5} -1)\, q^{-(3+4/2 -1)}\hat{T}(3,1) = (q-q^{-4})\, \hat{T}(3,1).
\label{ex22}
\end{eqnarray}

In \rref{ex22} the $5$ terms corresponding to positive reroutings at the points $P'$, $R'$, 
$T'$, $S'$, and $Q'$, are given in that order. $\beta_{R'}, \beta_{T'}, \beta_{S'},\beta_{Q'}$ 
denote the integer starting points of the corresponding rerouted paths (the $P'$ rerouting is at the 
origin)  i.e. $\beta_{R'} = (1,0)$, $\beta_{T'} = (2,0)$, $\beta_{S'} = (2,-1)$, $\beta_{Q'} = (3,0) $ 
(see Figures \rref{combined-exampleIe} and \rref{combined-exampleId}). 

In the first equality of \rref{ex22} the signed area equation \rref{s3}, was used to relate four 
rerouting terms, corresponding to  $R', T', S'$ and $Q'$, to the first rerouting term at the 
origin $P'=O$, i.e. $\hat{T}(p_1{O}p_2)$. The  regularization rule as in Figure \ref{combined-exampleId} 
justifies the
intersection number $(q -1)$ for the rerouting term corresponding
to $S'$, so that it indeed has the same intersection number as the other
intersection points.  The adjustment factor from the missing rule appears as the multiplier 
$q^{-5}$ in the 5th term corresponding to $Q'$. The passage from  $\hat{T}(p_1{O}p_2)$ to $\hat{T}(3,1)$ 
in the last line comes from applying Pick's rule to the rerouting at the origin (the first diagram in Figure 
\ref{5reroutings}).

\section{Conclusions}\label{concl}
In this model of $2+1$ quantum gravity we have considered Wilson variables for a 
large class of loops on a torus, related by area phases. This class of loops 
consists of straight paths (i.e. straight in $\IR^2$) and crooked paths (i.e. paths 
that arise from the intersection of straight paths, and rerouting along one of them) and 
the paths they generate through their intersections, expressed as Goldman brackets, or 
commutators. We have substantially clarified the nature of these variables for crooked paths, and 
have achieved a much fuller description of the refined Goldman bracket for this larger class of 
loops. 
%In doing so, we have used the concepts of integer points and relative phases 
%for a crooked rerouting.

We have significantly enriched our understanding of the quantum nature of intersections 
(expressed as commutators) of both straight and crooked paths. We have described some new 
features of the theory of intersecting loops on a torus, and given three precise rules to be applied at 
intersections of both straight and crooked paths. These intersection rules guarantee the reproduction of the 
corresponding straight - straight path result. We have presented two concrete examples of 
combinations of different rules

Our methods are a substantial step forwards towards our ultimate goal of defining the refined bracket \rref{qgold} 
which closes on a suitable class of straight and crooked paths. 

There are a number of other questions which should be addressed in a fully consistent intersection 
and rerouting theory based on the refined bracket: 

\begin{itemize}
\item By means of the V-intersection rule of Section \ref{sub41}, we have found a way of dealing with the 
two different types of quantum intersection numbers which naturally appear in the two quantizations \rref{qgb1}
and \rref{qgold}. These have different properties (in particular, the intersection number in the direct 
bracket \rref{qgb1} is symmetric under the interchange $q \leftrightarrow q^{-1}$, whereas in the the 
refined bracket \rref{qgold} it is not).

\item In the full quantum intersection theory that is emerging we must prove the Jacobi identity for the 
refined bracket, extended to a wider class of paths as in Section \ref{sec4}.

\item A full mathematical formalism needs to be developed for the quantum intersection theory described here.

\item Although our techniques are specific to the torus, since we are using paths in the covering space, the 
underlying principles of intersections, reroutings, and area phases should apply to other spaces as well, including 
general genus punctured surfaces and possibly orbifolds coming from a discrete group acting on  ${\IR}^2$.

\item There are fascinating hints of an interplay with knot theory. The rerouting of loops at intersections 
is obviously reminiscent of a skein relation, as suggested also by Turaev's approach  \cite{Turaev}
to quantizing the Goldman bracket, which uses knots in the thickened surface and skein modules. Work by 
Gukov \cite{guk} involves knot complements, i.e. 3D manifolds whose boundary is a torus, for which the cycles 
satisfy a q-commutation relation, exactly like \rref{fund2}.

\item In future work we plan to explore in more detail the link with $2$ - dimensional
holonomy \cite{FMP}, as discussed at the end of Section 2, and gain further
understanding of the elegant quantum geometry that emerges through the quantized
Goldman bracket, relating it e.g. to noncommutative geometry \cite{con}, or the BTZ black hole
\cite{vazwit}.
   
\end{itemize}

\section*{Acknowledgements}

This work was supported by the Istituto Nazionale di Fisica Nucleare (INFN) of
Italy, Iniziativa Specifica MI12, the Italian Ministero dell' Universit\`a e
della Ricerca Scientifica e Tecnologica (MIUR), and the project New Geometry and Topology,
PTDC/MAT/101503/2008, financed by the {\em Funda\c{c}\~{a}o para a Ci\^{e}ncia e
a Tecnologia} (FCT) and cofinanced by the European Community fund FEDER.



\bibliographystyle{my-h-elsevier}

\end{document}